\documentclass[twocolumn,english,superscriptaddress,citeautoscript,showpacs,preprintnumbers,amsmath,amssymb,prd,floatfix,nofootinbib]{revtex4-1}
\pdfoutput=1
\usepackage{amsmath, amsthm, amssymb, graphicx,bm,amstext,subfigure, mathdots}
\usepackage{hyperref}
\usepackage{color}

\begin{document}
\title{Vacuum fluctuation, microcyclic ``universes" and the cosmological constant problem}
\author{Qingdi Wang}
\author{William G. Unruh}
\affiliation{Department of Physics and Astronomy, 
The University of British Columbia,
Vancouver, Canada V6T 1Z1}
\begin{abstract}
We point out that the standard formulation of the cosmological constant problem itself is problematic since it is trying to apply the very large scale homogeneous cosmological model to very small (Planck) scale phenomenon. At small scales, both the spacetime and the vacuum stress energy are highly inhomogeneous and wildly fluctuating. This is a version of Wheeler's ``spacetime foam". We show that this ``foamy" structure would produce a large positive contribution to the average macroscopic spatial curvature of the Universe. In order to cancel this contribution to match the observation, the usually defined effective cosmological constant $\lambda_{\mathrm{eff}}=\lambda_B+8\pi G\langle\rho\rangle$ has to take a large negative value. The spacetime dynamics sourced by this large negative $\lambda_{\mathrm{eff}}$ would be similar to the cyclic model of the universe in the sense that at small scales every point in space is a ``micro-cyclic universe" which is following an eternal series of oscillations between expansions and contractions. Moreover, if the bare cosmological constant $\lambda_B$ is dominant, the size of each ``micro-universe" would increase a tiny bit at a slowly accelerating rate during each micro-cycle of the oscillation due to the weak parametric resonance effect produced by the fluctuations of the quantum vacuum stress energy tensor. In this way, the large cosmological constant generated at small scales is hidden at observable scale and no fine-tuning of $\lambda_B$ to the accuracy of $10^{-122}$ is needed. This at least resolves the old cosmological constant problem and suggests that it is the quantum vacuum fluctuations serve as the dark energy which is accelerating the expansion of our Universe.
\end{abstract}
\maketitle

\section{Introduction}\label{problem}
The cosmological constant problem is widely regarded as one of the major obstacles to further progress in fundamental physics (e.g., see \cite{witten, RevModPhys.61.1, Kolb:EU90, Dolgov:1997za, citeulike:430764}). This problem arises at the intersection between quantum mechanics and general relativity. Basically, the uncertainty principle of quantum mechanics predicts that the quantum fields vacuum possesses a huge amount of energy. Then the equivalence principle of general relativity requires that this huge energy must gravitate to produce a large gravitational effect. However, this supposed large gravitational effect is not observed. The discrepancy between theory and experiment is as high as $122$ orders of magnitude depending on the high energy cutoff and other factors. This is undoubtedly the largest discrepancy in all of science and is thus called the ``worst theoretical prediction in the history of physics".

Most proposed solutions to the cosmological constant problem are either trying to modify quantum mechanics in some way to make the vacuum energy small, trying to modify general relativity in some way to make the huge energy not gravitate or even pleading the anthropic principle. Unlike these proposals in literature, in \cite{PhysRevD.95.103504, PhysRevD.98.063506} we made a proposal for addressing this problem without modifying either quantum mechanics or general relativity.

In our proposal, the vacuum energy is still large as predicted by quantum mechanics and this huge energy does gravitate according to general relativity. However, the density of energy in the quantum vacuum is not a constant as usually assumed but is constantly fluctuating with its magnitude of fluctuation as big as its expectation value. As a result, the gravitational effect of the quantum vacuum would be different from what people previously thought. The resulting spacetime sourced by quantum vacuum would also be fluctuating and becomes highly inhomogeneous. Thus the gravitational effect produced by the huge vacuum stress energy is still huge, but is confined to small scales where each spatial point oscillates between expansion and contraction with different phase from neighboring spatial points. The expansion and the contraction almost cancel in this effect except the expansion wins out a little bit due to the weak parametric resonance effect produced by the vacuum fluctuations. This tiny net expansion accumulates on cosmological scale, gives the observed slowly accelerating expansion of the Universe.

In \cite{PhysRevD.95.103504, PhysRevD.98.063506} the calculations are performed for a highly simplified metric:
\begin{equation}\label{old metric}
ds^2=-dt^2+a^2(t, \mathbf{x})\left(dx^2+dy^2+dz^2\right).
\end{equation}
However, quantum fluctuations posses rich structures. One does not expect the spacetime to take the above simple form. In this paper, we start our calculations from the most general metric
\begin{equation}\label{new metric}
ds^2=-N^2dt^2+h_{ab}(dx^a+N^adt)(dx^b+N^bdt),
\end{equation}
where the spatial metric $h_{ab}$ depends on both time and space. This metric defines a $3+1$ decomposition of the spacetime with spatial slices $\{t=Constants\}$. The lapse function $N$ and the shift vector $N^a$ are not dynamical quantities. Starting from an initial slice $\Sigma_0$ defined by $t=0$, different choices of $N$ and $N^a$ give different spatial slices but they describe the physically equivalent spacetime. We are going to adopt the most convenient choice $N=1$, $N^a=0$, i.e. the Gaussian normal coordinates \eqref{Gaussian normal coordinate} to study the dynamical evolution of the spacetime. 

It turns out that this is not a straightforward generalization of our old papers \cite{PhysRevD.95.103504, PhysRevD.98.063506}. A new scenario which is different in some crucial ways is necessary.

In the old scenario presented in \cite{PhysRevD.95.103504, PhysRevD.98.063506}, the bare cosmological constant $\lambda_B$ in the Einstein field equations is set to zero for the highly simplified metric \eqref{old metric}. However, in the new scenario we are going to present in this paper, $\lambda_B$ can not be set to zero for the general metric \eqref{new metric}. The metric \eqref{new metric} allows more freedoms for the spacetime fluctuations and we find that these fluctuations would give a large positive contribution to the average macroscopic spatial curvature of the Universe. In order to match the observed small spatial curvature, one has to take $\lambda_B$ to large negative values to cancel it. It will turn out that $\lambda_B$ does not need to be carefully chosen (not fine-tuned).

Therefore, we are going to keep the bare cosmological constant $\lambda_B$ from the beginning and take it to have large negative values. Unlike the old scenario where we take the high energy cutoff $\Lambda$ to large positive values to obtain the alternatively expanding and contracting spacetime at small scales, we can obtain the similar picture for a fixed $\Lambda$ if its value is small compared to $\sqrt{|\lambda_B|}$. There is also the weak parametric resonance effect produced by the vacuum stress energy tensor fluctuations which may drive the accelerating expansion of the Universe. The new scenario also avoids some of the shortcomings of the old one. This will be explained in Sec.\ref{Advantages of the new scenario}.

This paper is organized as follows: in Sec.\ref{Problems of the cosmological constant problem} we review the standard formulation of the cosmological constant problem and point out where is wrong with this formulation; in Sec.\ref{approach} we explain the approach we are going to use in this paper; in Sec.\ref{The spatial curvature and the cosmological constant} we study the effect of small scale spacetime fluctuation on the averaged spatial curvature; in Sec.\ref{The evolution equation} we derive the key evolution equation of this paper; in sec.\ref{The effect of spacetime fluctuation} we study the effect of small scale spacetime fluctuation; in Sec.\ref{The effect of vacuum stress energy tensor fluctuation} we study the effect of vacuum stress energy tensor fluctuation; in Sec.\ref{discussions} we discuss the issue of the definition of vacuum state in our wildly fluctuating spacetime, the validity of our classical treatment of the spacetime evolution, the issue of the singularities appeared in this scenario, explain how the new scenario avoids a couple of shortcomings of the old one and  list some open questions arose from our model; in Sec.\ref{Summary and conclusion} we summarize the approach and the result of this paper and make the conclusions.

\section{Problems of the cosmological constant problem}\label{Problems of the cosmological constant problem}
The cosmological constant problem arises when one tries to put quantum mechanics and general relativity together to study the gravitational property of quantum vacuum. Since we do not have a satisfactory quantum theory of gravity yet, the usual assumption is the semiclassical Einstein equations
\begin{equation}\label{Einstein equation usual form}
G_{ab}+\lambda_B g_{ab}=8\pi G \langle T_{ab}\rangle,
\end{equation}
where $\lambda_B$ is the bare cosmological constant and the source of gravity is the expectation value of the quantum vacuum stress energy tensor. Vacuum is assumed to be Lorentz invariant and thus the expectation value $\langle T_{ab}\rangle$ is supposed to satisfy the vacuum equation of state
\begin{equation}\label{vacuum equation of state}
\langle T_{ab}\rangle=-\langle\rho\rangle g_{ab},
\end{equation}
where the expectation value of the vacuum energy density $\langle\rho\rangle$ has to be a constant due to the conservation of the stress energy tensor $\nabla^aT_{ab}=0$.

Then the gravitational effect of the vacuum would be equivalent to a cosmological constant that the Einstein equations \eqref{Einstein equation usual form} can be written as
\begin{equation}\label{vacuum einstein equation}
G_{ab}+\lambda_{\text{eff}} g_{ab}=0,
\end{equation}
where the effective cosmological constant $\lambda_{\text{eff}}$ is defined by
\begin{equation}\label{contributions to effective cc new new}
\lambda_{\text{eff}}=\lambda_B+8\pi G\langle\rho\rangle.
\end{equation}

In principle, all known and unknown fundamental matter fields would contribute to $\langle\rho\rangle$. The dominant contribution to $\langle\rho\rangle$ comes from the quantum zero-point energies of these fundamental fields. Without the knowledge of all fundamental fields, it is impossible to determine the exact value of $\langle\rho\rangle$. However, the standard effective field theory arguments predict that, in general, $\langle\rho\rangle$ takes the form
\begin{equation}
\langle\rho\rangle\sim\Lambda^4,
\end{equation}
if we trust our theory up to a certain high energy cutoff $\Lambda$. This result could have been guessed by dimensional analysis and the numerical constants which have been neglected will depend on the precise knowledge of the fundamental fields under consideration \cite{Carroll:2000fy}. The exact value of the cutoff $\Lambda$ is also not known. If it is taken to be the Planck energy, i.e. $\Lambda=1$, we would have
\begin{equation}\label{theorectical value}
\langle\rho\rangle\sim 1,
\end{equation}
where Planck units has been used for convenience.

One crucial assumption in the formulation of the cosmological constant problem is that the spacetime is homogeneous and isotropic, i.e. one assumes the standard FLRW metric of cosmology:
\begin{equation}\label{FLRW}
ds^2=-dt^2+a^2(t)(dx^2+dy^2+dz^2).
\end{equation}
The solution to the Einstein equation \eqref{vacuum einstein equation} under this metric is
\begin{equation}
a(t)=a(0)e^{Ht},
\end{equation}
where
\begin{equation}\label{relation}
H=\frac{\dot{a}}{a}=\pm \sqrt{\frac{\lambda_{\mathrm{eff}}}{3}}.
\end{equation}
Depending on the initial conditions, the Hubble rate $H$ can be either positive or negative. The ``$+$" sign represents an accelerated expanding universe while the ``$-$" sign represents an accelerated contracting universe. The value of the effective cosmological constant $\lambda_{\mathrm{eff}}$ can be determined by the observed rate of the accelerating expansion of the Universe $H$ from the relation \eqref{relation} that
\begin{equation}\label{observed value}
\lambda_{\mathrm{eff}}=3H^2=5.6\times 10^{-122},
\end{equation}
where Planck units has been used for convenience.

Therefore, the observed value of the effective cosmological constant $\lambda_{\mathrm{eff}}$ given by \eqref{observed value} is different from the theoretical value of the vacuum energy density $\langle\rho\rangle$ given by \eqref{theorectical value} by $122$ orders of magnitude.  Thus, according to \eqref{contributions to effective cc new new}, one has to fine-tune $\lambda_B$ to a precision of 122 decimal places to cancel $\langle\rho\rangle$ to match the observations. This problem of extreme fine-tuning is the so called cosmological constant problem \cite{RevModPhys.61.1}.

Such a large discrepancy between theory and observation implies that there must be something wrong in the above standard formulation of the cosmological constant problem. 

We have proposed that what is wrong is the neglection of the quantum vacuum and the spacetime fluctuaions in our previous paper \cite{PhysRevD.95.103504}. In the following we give a more comprehensive argument about this point.

The problem of the standard formulation comes from trying to apply the very large scale cosmological model to very small (Planck) scale phenomenon. Since the large contribution to the cosmological constant from the vacuum is generated by very small (Planck) scale quantum fluctuations, one should also look for answers directly at that scale \cite{Carlip:2018zsk}. There is no reason to expect the cosmological FLRW metric \eqref{FLRW} is still applicable at such small scales. In fact, the FLRW metric assumes homogeneous matter distribution and spacetime, but at small scales both the matter field vacuum and the spacetime are highly inhomogeneous and wildly fluctuating.

First, the vacuum energy density $\rho$ is not a constant because the vacuum is not an eigenstate of the energy density operator $T_{00}$, although it is an eigenstate of the Hamiltonian $\mathcal{H}=\int dx^3 T_{00}$, which is an integral of $T_{00}$ over the whole space. I.e., $\rho$ is not a constant because $T_{00}$ does not commute with $\mathcal{H}$. So the average energy density over a relatively large length scale ($\gg 1/\Lambda$) is nearly constant, but, at small length scales ($\sim 1/\Lambda$), $\rho$ can not be a constant, it is always fluctuating. In fact, the magnitude of the fluctuation is as large as its expectation value \cite{PhysRevD.95.103504}
\begin{equation}
\Delta\rho\sim\langle \rho\rangle.
\end{equation}
More detailed analysis shows that these fluctuations are highly inhomogeneous at small scales \cite{PhysRevD.95.103504}. In general, since the vacuum state is not an eigenstate of the stress energy tensor operator, the whole stress energy tensor would be violently fluctuating and highly inhomogeneous. 

Then the resulting spacetime sourced by such wlidly fluctuating and highly inhomogeneous vacuum energy is not homogeneous. Moreover, besides these ``passive" fluctuations driven by the fluctuations of the matter field vacuum stress energy tensor, the spacetime also experiences ``active" flucutaions due to the quantum nature of gravity itself. This was already anticipated by John Wheeler \cite{Wheeler:1957mu, misner1973gravitation} in 1955 that over sufficiently small distances and sufficiently small brief intervals of time, the ``very geometry of spacetime fluctuates". The spacetime would have a foamy, jittery nature and would consist of many small, ever-changing, regions. This picture of highly inhomogeneous fluctuating spacetime is called ``spacetime foam".

Therefore, we should not trust the standard formulation of the cosmological constant problem which is based on the homogeneous FLRW metric.

\section{Our approach}\label{approach}
One of the most important features of a quantum system is the quantum fluctuation due to the uncertainty principle. We have argued that the standard formulation of the cosmological constant problem missed the important fluctuations in the spacetime metric and in the matter fields vacuum stress energy tensor. In principle, one should use a quantum theory of gravity to study the effects of these fluctuations. But unfortunately, no satisfactory quantum theory of gravity exists yet.

For this reason, we are not trying to quantize gravity in this paper. Instead, we are still using the classical Einstein field equations
\begin{equation}\label{classical EFT}
G_{\mu\nu}+\lambda_Bg_{\mu\nu}=8\pi GT_{\mu\nu},
\end{equation}
where both the spacetime metric $g_{\mu\nu}$ and the matter fields stress energy tensor $T_{\mu\nu}$ are classical. In order to capture the essential feature of quantum fluctuations, both $g_{\mu\nu}$ and $T_{\mu\nu}$ are modeled as classical fluctuating fields. This treatment should at least reveal some quantum fluctuation feature of the future satisfactory quantum theory of gravity. This approximation is known as stochastic gravity \cite{Hu:2008rga}.

Technically, we will employ the initial value formulation of general relativity (see, e.g., \cite{Wald:1984rg, Alan:PDEinGR}) where the spacetime is decomposed as $3$-dimensional spacelike hypersurfaces plus $1$-dimensional time to study the effects of the fluctuations in the spacetime metric and in the matter fields vacuum stress energy tensor.

Let $\Sigma_t$, parameterized by a time function $t$, be spacelike Cauchy surfaces and $n^a$ be the unit normal vector field to $\Sigma_t$. The spacetime metric $g_{ab}$, induces a spatial metric $h_{ab}$ on each $\Sigma_t$ by
\begin{equation}
h_{ab}=g_{ab}+n_an_b.
\end{equation}

Let $t^a$ be a vector field satisfying $t^a\nabla_at=1$ which representing the ``flow of time" throughout the spacetime. The lapse function $N$ and the shift vector $N^a$, with respect to $t^a$ are defined by
\begin{eqnarray}
&&N=-t^an_a,\\
&&N_a=h_{ab}t^b.
\end{eqnarray}
Then the four-dimensional metric $g_{ab}$ can be written as
\begin{equation}\label{most general metric}
ds^2=-N^2dt^2+h_{ab}(dx^a+N^adt)(dx^b+N^bdt).
\end{equation}

The extrinsic curvature $K_{ab}$ of the hypersurface $\Sigma_t$ is related to the time derivative $\dot{h}_{ab}$ of the spatial metric by
\begin{equation}\label{kab geometric meaning}
K_{ab}=\frac{1}{2}N^{-1}\left(\dot{h}_{ab}-D_aN_b-D_bN_a\right),
\end{equation}
where $D_a$ is the derivative operator on $\Sigma_t$ associated with $h_{ab}$.

In the initial value formulation, the Einstein field equations are equivalent to six equations for the time evolution of the extrinsic curvature
\begin{eqnarray}\label{kabdian}
\dot{K}_{ab}=&&-N\bigg[-\lambda_Bh_{ab}+R^{(3)}_{ab}+KK_{ab}-2K_{ac}K^c_b  \nonumber\\
&&-4\pi G\rho h_{ab}-8\pi G\left(T_{ab}-\frac{1}{2}h_{ab}\text{tr}T\right)\bigg]\\
&&+D_aD_bN+N^cD_cK_{ab}+K_{ac}D_bN^c+K_{cb}D_aN^c, \nonumber
\end{eqnarray}
plus the Hamiltonian and momentum constraint equations
\begin{eqnarray}
R^{(3)}-K_{ab}K^{ab}+K^2&=&16\pi G\rho+2\lambda_B,  \label{Hamiltonian constraint}\\
D^aK_{ab}-D_bK&=&-8\pi G J_b,
\end{eqnarray}
where $R^{(3)}_{ab}$ is the $3$-dimensional Ricci tensor of $\Sigma$, $R^{(3)}=h^{ab}R^{(3)}_{ab}$ is the $3$-dimensional Ricci scalar of $\Sigma$, $K=h^{ab}K_{ab}$ is the mean curvature of $\Sigma$, $\rho=T_{ab}n^an^b$, $J_b=-h_b^cT_{ca}n^a$ and $\text{tr}T=h^{ab}T_{ab}$ are the energy density, the energy flux and the spatial trace of the matter fields, respectively.

\section{The spatial curvature fluctuation}\label{The spatial curvature and the cosmological constant}
One immediate consequence of the spacetime foam picture is that the spatial curvature $R^{(3)}$ of the Cauchy surface $\Sigma_t$ at each point would be large and fluctuating. However, the observed average spatial curvature of the Universe is very small (flat with only a 0.4 percent margin of error). So one natural question is: can the large curvature at small scales averages to small value macroscopically?

For a given spacetime, there are infinite ways to perform the $3+1$ decomposition. Different splitting leads to different spatial curvature $R^{(3)}$ of $\Sigma_t$, i.e., this question highly depends on how we choose the spatial slice $\Sigma_t$. For example, even for the flat Minkowski spacetime, there are infinite ways to choose $\Sigma_t$ such that $R^{(3)}$ is not zero. Therefore, a more precise description about this question is: for our fluctuating spacetime, can we find a $3+1$ decomposition such that for each $\Sigma_t$, the average spatial curvature $\langle R^{(3)}\rangle$ approaches zero?


To answer this question, let us start from an arbitrary Cauchy surface $\tilde{\Sigma}$. Taking the spatial average of \eqref{Hamiltonian constraint} over $\tilde{\Sigma}$ we get the average spatial curvature
\begin{equation}\label{average spatial curvature}
\langle R^{(3)}\rangle_{\tilde{\Sigma}}=2\lambda_{\mathrm{eff}}+\langle K_{ab}K^{ab}-K^2\rangle_{\tilde{\Sigma}},
\end{equation}
where $\lambda_{\mathrm{eff}}=\lambda_B+8\pi G\langle\rho\rangle$ is the effective cosmological constant in the standard formulation of the cosmological constant problem defined by \eqref{contributions to effective cc new new}.

The term $K_{ab}K^{ab}-K^2$ can be expanded as
\begin{eqnarray}\label{kab-ksquare}
&&K_{ab}K^{ab}-K^2 \nonumber\\
=&&\left(h^{ac}h^{bd}-h^{ab}h^{cd}\right)K_{ab}K_{cd} \\
=&&\sum_{i\neq j\neq k}M_kK_{ij}^2+\sum_{\{i, j\}\neq\{k, l\}}\left(h^{ik}h^{jl}-h^{ij}h^{kl}\right)K_{ij}K_{kl}, \nonumber
\end{eqnarray}
where
\begin{equation}
M_k=h^{ii}h^{jj}-\left(h^{ij}\right)^2, \quad k\neq i\neq j,
\end{equation}
is the determinant of the submatrix formed by deleting the $k$th row and $k$th column of the $3\times 3$ symmetric matrix $h^{ab}$, i.e. it is the $k$th principal minor of $h^{ab}$. Since by definition the metric matrix $h^{ab}$ is positive definite, we have $M_k>0$.


Since general relativity is time reversal invariant that for every expanding solution there is a corresponding contracting solution, i.e., if $(h_{ab}, K_{ab})$ is allowed initial data on $\tilde{\Sigma}$, so is $(h_{ab}, -K_{ab})$ \cite{Carlip:2018zsk}. Thus, for $\{i, j\}\neq\{k,l\}$, the following four pairs of components
\begin{equation}
(K_{ij}, K_{kl}),\,(K_{ij}, -K_{kl}),\,(-K_{ij}, K_{kl}),\,(-K_{ij}, -K_{kl}) \nonumber
\end{equation}
are equally likely to happen for a large collection of possible choices of $\tilde{\Sigma}$. Then because in general, there is no particular relationship between the components of the extrinsic curvature, we have, for the second term in \eqref{kab-ksquare}, the above four cases would statistically cancel each other that the macroscopic spatial average over $\tilde{\Sigma}$:
\begin{equation}\label{zero macroscpic average}
\left\langle\left(h^{ik}h^{jl}-h^{ij}h^{kl}\right)K_{ij}K_{kl}\right\rangle_{\tilde{\Sigma}}=0, \quad \{i, j\}\neq\{k, l\}.
\end{equation}
Then taking the macroscopic spatial average on both sides of \eqref{kab-ksquare} we obtain
\begin{equation}\label{kab average}
\langle K_{ab}K^{ab}-K^2\rangle_{\tilde{\Sigma}}=\sum_{i\neq j\neq k}\langle M_kK_{ij}^2\rangle>0
\end{equation}
for a large collection of possible choices of $\tilde{\Sigma}$. Note that the macroscopic average does not require a very large volume: a cubic centimeter contains some $10^{100}$ Planck-size regions.

The term $\langle M_kK_{ij}^2\rangle$ in \eqref{kab average} is very large in the wildly fluctuating spacetime. It gives a large positive contribution to the average spatial curvature $\langle R^{(3)}\rangle_{\tilde{\Sigma}}$ through \eqref{average spatial curvature}. This implies that, for a large collection of possible choices of $\tilde{\Sigma}$, $\lambda_{\mathrm{eff}}$ has to take large negative values to make $\langle R^{(3)}\rangle_{\tilde{\Sigma}}$ small to match the observation\footnote{It is interesting to apply the same argument to the old highly simplified metric \eqref{old metric} we used in \cite{PhysRevD.95.103504}. In this case, the only nonzero components of the extrinsic curvature are $K_{11}=K_{22}=K_{33}=a\dot{a}$ so that we have
\begin{equation}
K_{ab}K^{ab}-K^2=-6\left(\frac{\dot{a}}{a}\right)^2.
\end{equation}
Then the spatial curvature
\begin{eqnarray}
R^{(3)}&=&2\left(\lambda_B+8\pi G\rho-3\left(\frac{\dot{a}}{a}\right)^2\right)\label{1}\\
&=&-\frac{2}{a^2}\left(\nabla\left(\frac{\nabla a}{a}\right)+\frac{\nabla^2a}{a}\right), \label{2}
\end{eqnarray}
where we have used the $00$ component of the Einstein equation $G_{00}=0$ when deriving \eqref{2} from \eqref{1}. For a large collection of initial data on the hypersurface $t=0$, since there is not a special spatial direction, the average of the gradient terms of $a$ in \eqref{2} over the hypersurface $t=0$ should approaches zero. So we would have
\begin{equation}
\langle R^{(3)}\rangle=2\left(\lambda_{\mathrm{eff}}-3\left\langle\left(\frac{\dot{a}}{a}\right)^2\right\rangle\right)\approx 0,
\end{equation}
and thus
\begin{equation}
\lambda_{\mathrm{eff}}\approx 3\left\langle\left(\frac{\dot{a}}{a}\right)\right\rangle^2>0.
\end{equation}
In \cite{PhysRevD.95.103504}, $\lambda_B$ is set to zero, so that the above condition requires a positive vacuum energy density
\begin{equation}
8\pi G\langle\rho\rangle\approx 3\left\langle\left(\frac{\dot{a}}{a}\right)\right\rangle^2>0
\end{equation}
to make sure $\langle R^{(3)}\rangle\approx 0$. 

So in this case we get opposite result for the sign of $\lambda_{\mathrm{eff}}$. In addition, $\langle\rho\rangle$ can be any sign in this paper since we can always adjust $\lambda_B$ to make $\langle R^{(3)}\rangle\approx 0$.}:
\begin{equation}\label{negative lambda eff}
\lambda_{\mathrm{eff}}\approx -\frac{1}{2}\langle K_{ab}K^{ab}-K^2\rangle_{\tilde{\Sigma}}<0.
\end{equation}

It seems that this leads to another fine-tuning problem: for the given Cauchy surface $\tilde{\Sigma}$, one has to fine-tune $\lambda_B$ to cancel the large term $8\pi G\langle\rho\rangle+\frac{1}{2}\langle K_{ab}K^{ab}-K^2\rangle_{\tilde{\Sigma}}$ to obtain a small $\langle R^{(3)}\rangle_{\tilde{\Sigma}}$. However, remember that our task is not to pick an arbitrary hypersurface and tune $\lambda_B$ to make its average spatial curvature small. Our task is to find a hypersurface whose average spatial curvature is small for a given $\lambda_B$. This can be done by the following procedure.

We construct a family of hypersurfaces $\tilde{\Sigma}_s$ by continuously deforming the 3-dimensional hypersurface $\tilde{\Sigma}$ in the given fluctuating 4-dimensional spacetime. The spatial averages $\langle K_{ab}K^{ab}-K^2\rangle_{\tilde{\Sigma}_s}$ would then change continuously from $\langle K_{ab}K^{ab}-K^2\rangle_{\tilde{\Sigma}}$ that it would lie in a range
\begin{equation}
\langle K_{ab}K^{ab}-K^2\rangle_{\tilde{\Sigma}_s}\in [\tilde{a}, \tilde{b}], \quad \tilde{a}, \tilde{b}>0.
\end{equation}


We set $\lambda_B$ to be in the range:
\begin{equation}\label{range}
\lambda_B\in [-\frac{\tilde{b}}{2}-8\pi G\langle\rho\rangle, \,-\frac{\tilde{a}}{2}-8\pi G\langle\rho\rangle].
\end{equation}
Then there exists a hypersurface $\Sigma_0$ in the family of hypersurfaces $\tilde{\Sigma}_s$ such that
\begin{equation}
\langle K_{ab}K^{ab}-K^2\rangle_{\Sigma_0}=-2\lambda_{\mathrm{eff}}.
\end{equation}
In this way, we find an initial hypersurface $\Sigma_0$ for which the average spatial curvature
\begin{equation}
\langle R^{(3)}\rangle_{\Sigma_0}=0.
\end{equation}
Note that there is no need to fine-tune $\lambda_B$ to make the average spatial curvature of $\Sigma_0$ to be zero since $\lambda_B$ can take any values in the range given by \eqref{range}.


So far we have found an initial Cauchy surface $\Sigma_0$ whose average spatial curvature is small. Next question is whether this feature is preserved dynamically. The evolution equation for $\langle R^{(3)}_{ab}\rangle$ is (see, e.g., \cite{Alan:PDEinGR})
\begin{eqnarray}\label{rab evolution}
\dot{\langle R^{(3)}_{ab}\rangle}=&&-\langle D^cD_c(NK_{ab})\rangle-\langle D_aD_b(NK)\rangle\nonumber\\
&&+\langle D^cD_a(NK_{cb})\rangle+\langle D^cD_b(NK_{ca})\rangle\\
&&+\langle N^cD_cR^{(3)}_{ab}\rangle+\langle R^{(3)}_{ac}D_bN^c\rangle+\langle R^{(3)}_{cb}D_aN^c\rangle \nonumber
\end{eqnarray}

Although the spacetime is fluctuating, it should have the same property everywhere. So that there is not a special spatial direction we should have the spatial averages
\begin{eqnarray}
\langle D_aK\rangle&=&0, \label{property 1}\\
\langle D_aK_{bc}\rangle&=&0,\label{property 2}\\
\langle D_aR^{(3)}\rangle&=&0,\label{property 3}\\
\langle D_cR^{(3)}_{ab}\rangle&=&0.\label{property 4}
\end{eqnarray}
We can also choose $N$, $N^a$ such that the spatial averages
\begin{eqnarray}
\langle N\rangle&=&1, \label{property 5}\\
\langle N^a\rangle&=&0,\label{property 6}\\
\langle D_aN\rangle&=&0,\label{property 7}\\
\langle D_aN^b\rangle&=&0.\label{property 8}
\end{eqnarray}
Then since there is no particular relationships between $N$ and $K$(or $K_{ab}$), $N^a$ and $R^{(3)}_{ab}$, all the terms on the right hand side of \eqref{rab evolution} should be zero that we obtain
\begin{equation}
\dot{\langle R^{(3)}_{ab}\rangle}=0.
\end{equation}
Therefore, if $\langle R^{(3)}_{ab}\rangle=0$ at the initial hypersurface $\Sigma_0$, it should still be zero afterwards.

The evolution equation for $\langle R^{(3)}\rangle$ can be obtained from \eqref{rab evolution} that
\begin{eqnarray}\label{average evolution of R}
\dot{\langle R^{(3)}\rangle}=&&-2\langle NK^{ab}R^{(3)}_{ab}\rangle-2\langle D^aD_a(NK)\rangle \nonumber\\
&&+2\langle D^aD^b(NK_{ab})\rangle+\langle N^aD_aR^{(3)}\rangle.
\end{eqnarray}
Following similar arguments we have the last three terms on the right side of \eqref{average evolution of R} are zero. As for the first term, since we have $\langle R^{(3)}_{ab}\rangle=0$ and there is no particular relationship between the extrinsic curvature $K_{ab}$, which is given by the time derivative of $h_{ab}$, and the Ricci curvature $R^{(3)}_{ab}$, which is given by the spatial derivatives of $h_{ab}$, we would also have $\langle NK^{ab}R^{(3)}_{ab}\rangle=0$. Thus we reach the result
\begin{equation}
\dot{\langle R^{(3)}\rangle}=0,
\end{equation}
i.e. the small averaged macroscopic spatial curvature $\langle R^{(3)}\rangle$ is preserved with time.

Therefore, we have found a spacetime foliation $\{\Sigma_t\}$ that, the average spatial curvature $\langle R^{(3)}\rangle_{\Sigma_t}$ approaches zero for each $\Sigma_t$.

\section{The evolution equation}\label{The evolution equation}
In the last section we have shown that the randomly fluctuating foamy structure of the spacetime would give a large positive contribution to the averaged spatial curvature. In order to obtain the observed small spatial curvature of the Universe, $\lambda_{\mathrm{eff}}$ has to take large negative values to cancel this contribution. Next we study the dynamics of the spacetime when $\lambda_{\mathrm{eff}}$ takes the required negative value.

To start, we set $\lambda_{\mathrm{eff}}$ to satisfy \eqref{negative lambda eff} for the given initial hypersurface $\Sigma_0$ and a set of randomly chosen initial data $(h_{ab}, K_{ab})$ on it.

Taking the trace of \eqref{kabdian} we can obtain the evolution equation for the mean curvature $K$:
\begin{eqnarray}\label{kabdian1}
\dot{K}=&&-N\left(-3\lambda_B+R^{(3)}+K^2-12\pi G\rho+4\pi G\text{tr}T\right) \nonumber\\
&&+D^aD_aN+N^aD_aK.
\end{eqnarray}
Combining the above equation \eqref{kabdian1} with the Hamiltonian constraint \eqref{Hamiltonian constraint} gives
\begin{eqnarray}\label{kabdian2}
\dot{K}=&&-N\left[-\lambda_B+K_{ab}K^{ab}+4\pi G\left(\rho+\text{tr}T\right)\right] \nonumber\\
&&+D^aD_aN+N^aD_aK.
\end{eqnarray}

It is useful to split the extrinsic curvature $K_{ab}$ into the trace free part $\sigma_{ab}$ and the trace part $K$:
\begin{equation}\label{shear definition}
K_{ab}=\sigma_{ab}+\frac{1}{3}Kh_{ab}.
\end{equation}
The trace free part $\sigma_{ab}$ is called the shear tensor, its time evolution can be obtained by combining \eqref{kabdian} and \eqref{kabdian1}:
\begin{eqnarray}\label{shear evolution}
\dot{\sigma}_{ab}=&&-N\bigg[\frac{1}{3}K\sigma_{ab}-2\sigma_{ac}\sigma^c_b+R^{(3)}_{ab}-\frac{1}{3}R^{(3)}h_{ab} \nonumber\\
&&-8\pi G\left(T_{ab}-\frac{1}{3}h_{ab}\text{tr}T\right)\bigg]+D_aD_bN-\frac{1}{3}h_{ab}D^cD_cN  \nonumber\\
&&+N^cD_c\sigma_{ab}+\sigma_{ac}D_bN^c+\sigma_{cb}D_aN^c.
\end{eqnarray}
Plugging \eqref{shear definition} into \eqref{kabdian2} gives
\begin{eqnarray}\label{Raychaudhuri}
\dot{K}=&&-N\left[-\lambda_B+\frac{1}{3}K^2+2\sigma^2+4\pi G\left(\rho+\text{tr}T\right)\right] \nonumber\\
&&+D^aD_aN+N^aD_aK,
\end{eqnarray}
where $\sigma^2=\frac{1}{2}\sigma_{ab}\sigma^{ab}$. From \eqref{kab geometric meaning} we have that the mean extrinsic curvature
\begin{equation}\label{K relations}
K=\frac{1}{N}\left(\frac{\dot{h}}{2h}-D_aN^a\right),
\end{equation}
where $h=\mathrm{det}(h_{ab})$ is the determinant of the spatial metric.

The lapse function $N$ and the shift vector $N^a$ are not dynamical quantities, solutions with different choices of $N$ and $N^a$ are physically equivalent so that $N$ and $N^a$ can be freely chosen. We first choose the shift vector $N^a=0$. Then the metric \eqref{most general metric} becomes
\begin{equation}\label{metric with lapse}
ds^2=-N^2dt^2+h_{ab}dx^adx^b.
\end{equation}
In this metric, \eqref{K relations} becomes
\begin{equation}\label{111}
K=\frac{1}{\sqrt{h}}\frac{d}{d\tau}\sqrt{h},
\end{equation}
where $d\tau=Ndt$ is the proper time of the Eulerian observer defined by $\mathbf{x}=Constant$. Note that $\tau$ is different for different Eulerian observers. Since $\sqrt{h}\,dx^1\wedge dx^2\wedge dx^3$ is the spatial volume element, \eqref{111} means that $K$ is the local volume expansion rate of the $3$-dimensional hypersurface $\Sigma_t$ observed by the Eulerian observer. 

We can define a new quantity---the local scale factor $a(t, \mathbf{x})$ by
\begin{equation}\label{scale factor}
h=a^6.
\end{equation}
It locally describes the relative ``size" of space measured by the Eulerian observer at each point, which is a generalization of the scale factor $a(t)$ in the usual homogeneous FLRW metric \eqref{FLRW}. The difference is that now $a$ is also spatial dependent to be able to describe the fluctuating spacetime. Then we would have $K=\frac{3}{a}\frac{da}{d\tau}$ and from \eqref{Raychaudhuri} we obtain the evolution equation for $a$ observed by the Eulerian observer:
\begin{equation}\label{evolution with N}
\frac{d^2a}{d\tau^2}+\frac{1}{3}\left(2\sigma^2-\lambda_B+4\pi G\left(\rho+\mathrm{tr}T\right)-\frac{D^aD_aN}{N}\right)a=0.
\end{equation}

The term $D^aD_aN/N$ in \eqref{evolution with N} comes from the lapse function $N$. It represents the effect of external force acting on the Eulerian observer. In fact, the Eulerian observer has acceleration $a_i=D_iN/N$ (see Eq.(3.17) in \cite{Gourgoulhon:2007ue}) which is tangent to the spatial slices $\Sigma_t$. There are external forces acting on the Eulerian observers to maintain their constant spatial positions. We should exclude the effect of these external forces on the evolution of $a$ so that $a$ purely describes the gravitational effect produced by the terms $\sigma^2$, $\lambda_B$ and $\rho+\mathrm{tr}T$. To do this, we need to choose $N$ to be spatially independent to exclude the effect of the external forces or at least to choose $N$ in such a way that the average 
\begin{equation}\label{N condition}
\left\langle \frac{D^aD_aN}{N}\right\rangle=0
\end{equation}
to make sure the average effect of these external forces zero.

The simplest choice is $N=1$, then the coordinate \eqref{metric with lapse} reduces to the Gaussian normal coordinate
\begin{equation}\label{Gaussian normal coordinate}
ds^2=-dt^2+h_{ab}dx^adx^b.
\end{equation}
Then the evolution equation \eqref{evolution with N} becomes
\begin{equation}\label{a evolution}
\ddot{a}+\Omega^2a=0,
\end{equation}
where
\begin{equation}
\Omega^2=\frac{1}{3}\left(2\sigma^2-\lambda_B+4\pi G\left(\rho+\text{tr}T\right)\right).
\end{equation}

For simplicity, we are going to use the coordinate \eqref{Gaussian normal coordinate} in the following sections. Other choices of $N$ describe physically equivalent spacetime. In addition, our analysis using the coordinate\eqref{Gaussian normal coordinate} will also apply to the coordinate \eqref{metric with lapse} since the only difference comes from the term $D^aD_aN/N$ whose average effect is zero.

For convenience, we rewrite $\Omega^2$ as
\begin{equation}\label{Omega square rewritten}
\Omega^2=\frac{1}{3}\left(2\sigma^2-\lambda'_{\mathrm{eff}}\right)+F,
\end{equation}
where $\lambda'_{\mathrm{eff}}$ and $F$ are defined by
\begin{equation}
\lambda'_{\mathrm{eff}}=\lambda_B-4\pi G\langle\rho+\text{tr}T\rangle
\end{equation}
and
\begin{equation}\label{F definition}
F=\frac{4\pi G}{3}\left(\rho+\text{tr}T-\left\langle\rho+\text{tr}T\right\rangle\right).
\end{equation}
Note that by definition we have the expectation value
\begin{equation}
\langle F\rangle=0.
\end{equation}

$\lambda'_{\mathrm{eff}}$ is related to $\lambda_{\mathrm{eff}}$ by
\begin{equation}
\lambda'_{\mathrm{eff}}=\lambda_{\mathrm{eff}}-4\pi G\langle 3\rho+\text{tr}T\rangle
\end{equation}
If the usual vacuum equation of state \eqref{vacuum equation of state} is assumed, we have
\begin{equation}
\lambda'_{\mathrm{eff}}=\lambda_{\mathrm{eff}}.
\end{equation}
Whether the vacuum equation of state \eqref{vacuum equation of state} is still valid at very small (Planck) scales is controversial. The standard formulation of the cosmological constant problem assumes that it is valid. Later it will be clear this is not important, whether \eqref{vacuum equation of state} is valid or not does not affect our conclusion.

Note that the evolution equation \eqref{a evolution} does not contain spatial derivatives of the metric. Thus it is an ordinary differential equation whose solution at each spatial point $\mathbf{x}$ depends only on the initial values $a(0, \mathbf{x})$, $\dot{a}(0, \mathbf{x})$ and the time evolution of $\Omega^2(t, \mathbf{x})$ at $\mathbf{x}$. The solution to $a(t, \mathbf{x})$ at different spatial points explicitly decouple with each other, although implicitly they are not independent since there are correlations between $\Omega^2(t, \mathbf{x})$ at different spatial points.

\section{The effect of spacetime fluctuation}\label{The effect of spacetime fluctuation}
In order to understand the physical mechanism better, especially the role played by the spacetime fluctuation, we first exclude the term $F$ in the evolution equation \eqref{a evolution} which represents the vacuum stress energy tensor fluctuation, i.e. use the expectation value of the vacuum stress energy tensor in the Einstein equation.

Note that excluding vacuum stress energy fluctuation does not exclude spacetime fluctuation. As explained at the end of Sec.\ref{Problems of the cosmological constant problem} that the vacuum stress energy tensor fluctuation drives the ``passive'' fluctuation of the spacetime. However,  The spacetime still ``actively" fluctuates at small scales due to the quantum nature of gravity itself.


When excluding $F$, the evolution equation \eqref{a evolution} becomes
\begin{equation}\label{a evolution no vacuum}
\ddot{a}+\frac{1}{3}\left(2\sigma^2-\lambda'_{\mathrm{eff}}\right)a=0.
\end{equation}
The evolution equation for the shear $\sigma^2$ in the above equation \eqref{a evolution no vacuum} can be obtained by taking time derivative of $\sigma^2$ and using \eqref{shear evolution}:
\begin{equation}\label{shear evolution simplified}
(\sigma^2)^\cdot=-2K\sigma^2-R^{(3)}_{ab}\sigma^{ab},
\end{equation}
where the fluctuation of the vacuum stress energy tensor has also been excluded in the calculation. 

If the vacuum equation of state \eqref{vacuum equation of state} is assumed, we would have $\lambda'_{\mathrm{eff}}=\lambda_{\mathrm{eff}}=\lambda_B+8\pi G\langle\rho\rangle<0$. In the following, we are going to study the spacetime dynamics given by the above coupled equations \eqref{a evolution no vacuum} and \eqref{shear evolution simplified} when $\lambda'_{\mathrm{eff}}<0$ \footnote{If \eqref{vacuum equation of state} is not valid at small scales, $\lambda'_{\mathrm{eff}}$ may not be negative. However, the most interesting case we are going to study in the next section \ref{The effect of vacuum stress energy tensor fluctuation} is when $\lambda_B$ is dominant over the matter fields vacuum stress energy fluctuation. In this case, $\lambda'_{\mathrm{eff}}$ has to be negative no matter \eqref{vacuum equation of state} is valid or not.}.

\begin{figure}
\includegraphics[scale=0.6]{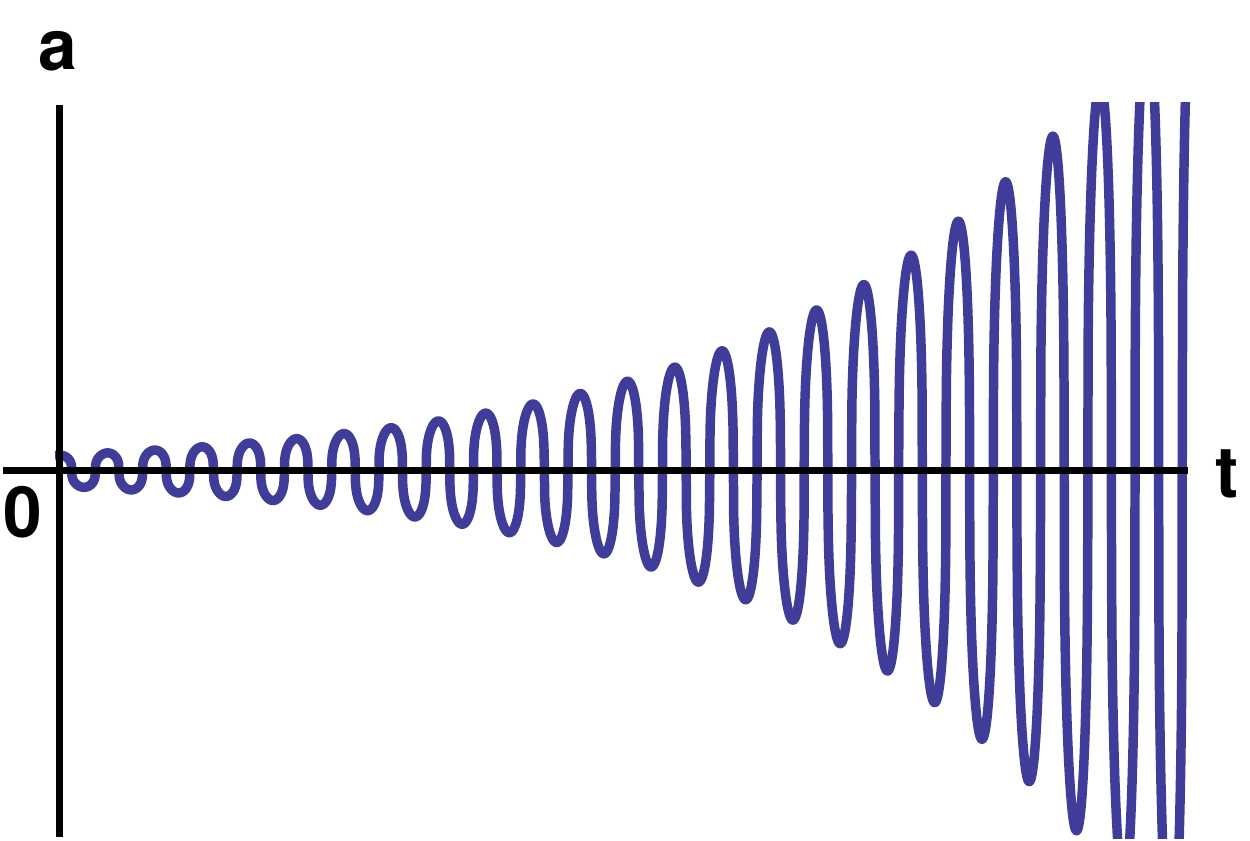}
\includegraphics[scale=0.6]{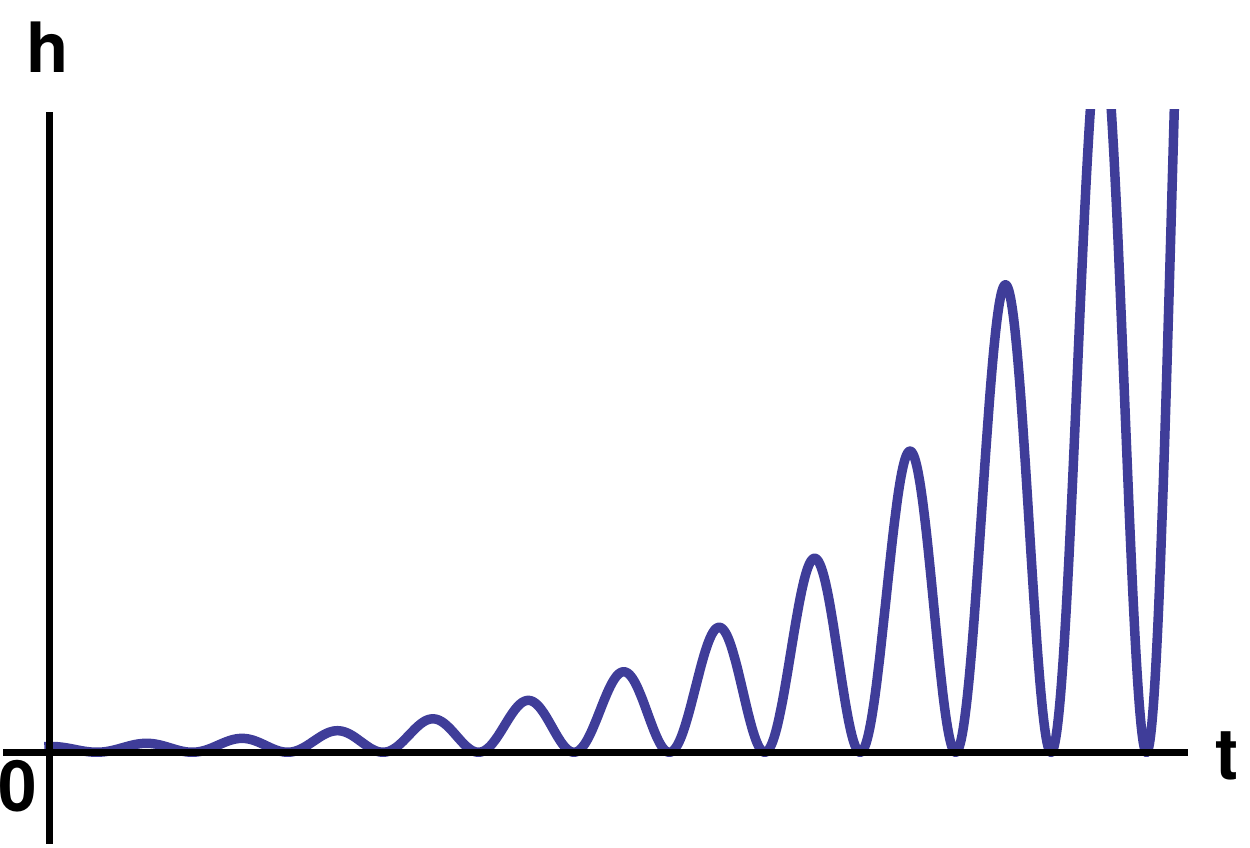}
\includegraphics[scale=0.63]{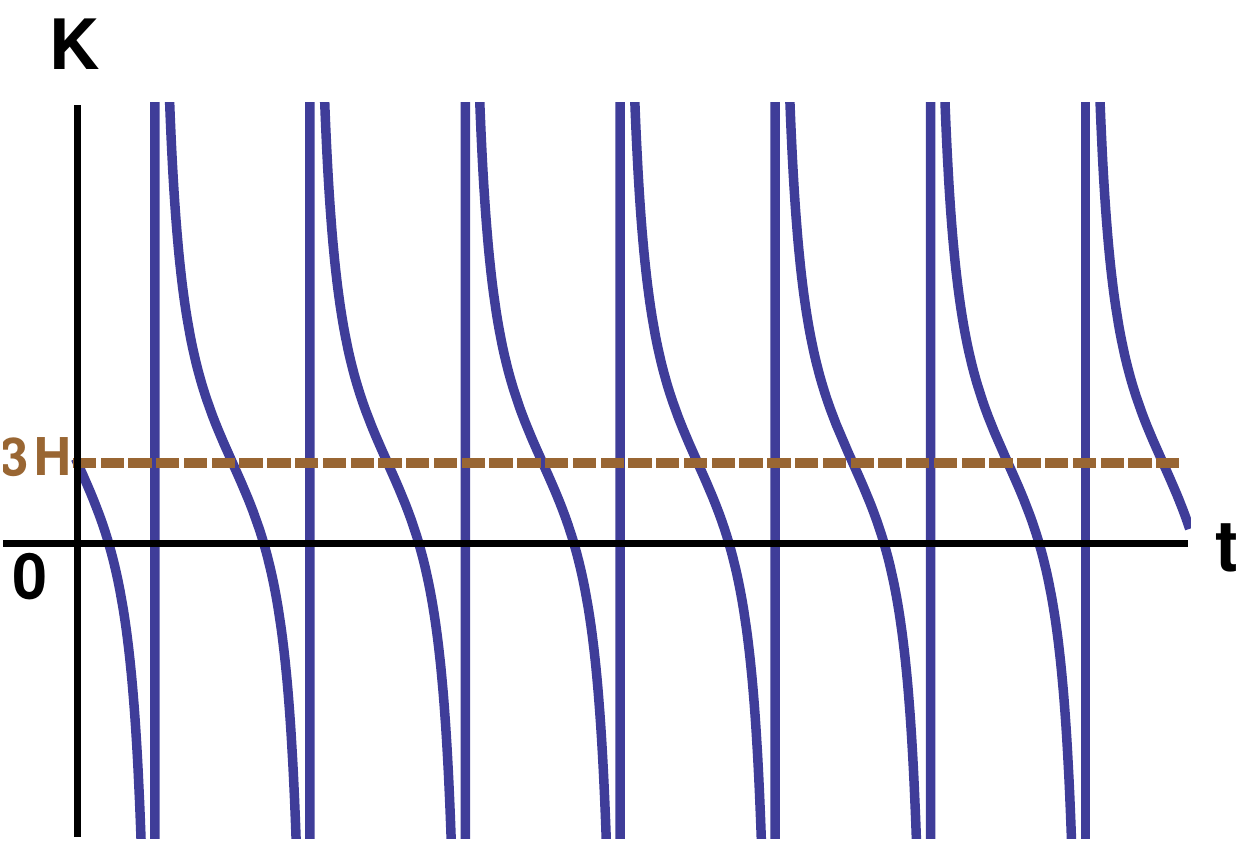}
\caption{\label{schematic plot}Schematic plots of the oscillations of the local scale factor $a$, the local determinant $h=a^6$ and the local average expansion rate $K=3\frac{\dot{a}}{a}$. As $a$ goes across $0$, $h$ decreases continuously to $0$ and then increases back to positive values, $K$ jumps discontinuously from $-\infty$ to $+\infty$. The amplitude of $a$ grows exponentially with a tiny exponent $H=\alpha\Lambda e^{-\beta\frac{\sqrt{-\lambda_B}}{\Lambda}}$ (Eq.\eqref{dependence of H on Lambda}) which gives the slowly accelerating expansion of $h$ and small average value $3H$ of $K$.}
\end{figure}

Since $\sigma^2>0$ and $-\lambda'_{\mathrm{eff}}>0$ that $\Omega^2=\frac{1}{3}\left(2\sigma^2-\lambda_{\mathrm{eff}}\right)$ must be positive, \eqref{a evolution no vacuum} describes an oscillator with varying frequency. Thus the solution for $a$ must be oscillating around $0$. Correspondingly, the local volume expansion rate $K=3\frac{\dot{a}}{a}$ would also oscillate. It ranges from $-\infty$ to $+\infty$. $K>0$ represents expansion while $K<0$ represents contraction. It jumps discontinuously from $-\infty$ to $+\infty$ each time when $a$ goes across $0$. In this process, the determinant $h=a^6\geq 0$ decreases continuously to $0$ and then bounces back to positive values as $a$ crosses $0$ (see FIG. \ref{schematic plot}). Physically, this means, on average, the space locally collapses to zero size and then immediately bounces back. It will then collapse and expand again and again, i.e., locally, the space is alternatively switching between expansion and contraction.

The oscillation behavior of $K$ would also lead to the oscillation of the shear $\sigma^2$. The average value of the second term $R^{(3)}_{ab}\sigma^{ab}$ in \eqref{shear evolution} is zero. So if we neglect this term, we would obtain that the average evolution of $\sigma^2$ roughly goes as
\begin{equation}
\bar{\sigma^2}(t, \mathbf{x})\sim \sigma^2(0, \mathbf{x}) e^{-2\int_0^{t} K(t', \mathbf{x}) dt'}.
\end{equation}
As $K>0$, i.e., as $a$ is moving away from its equilibrium point $a=0$,  $\bar{\sigma^2}$ is decreasing to a minimum until $|a|$ reaches maximum. As $K<0$, i.e., as $a$ is moving toward its equilibrium point $a=0$,  $\bar{\sigma^2}$ is increasing to a maximum until $|a|$ reaches $0$. In fact, since $K=3\frac{\dot{a}}{a}=\pm\infty$ at $a=0$, we have $\sigma^2=+\infty$ at $a=0$. 

\begin{figure}
\includegraphics[scale=0.24]{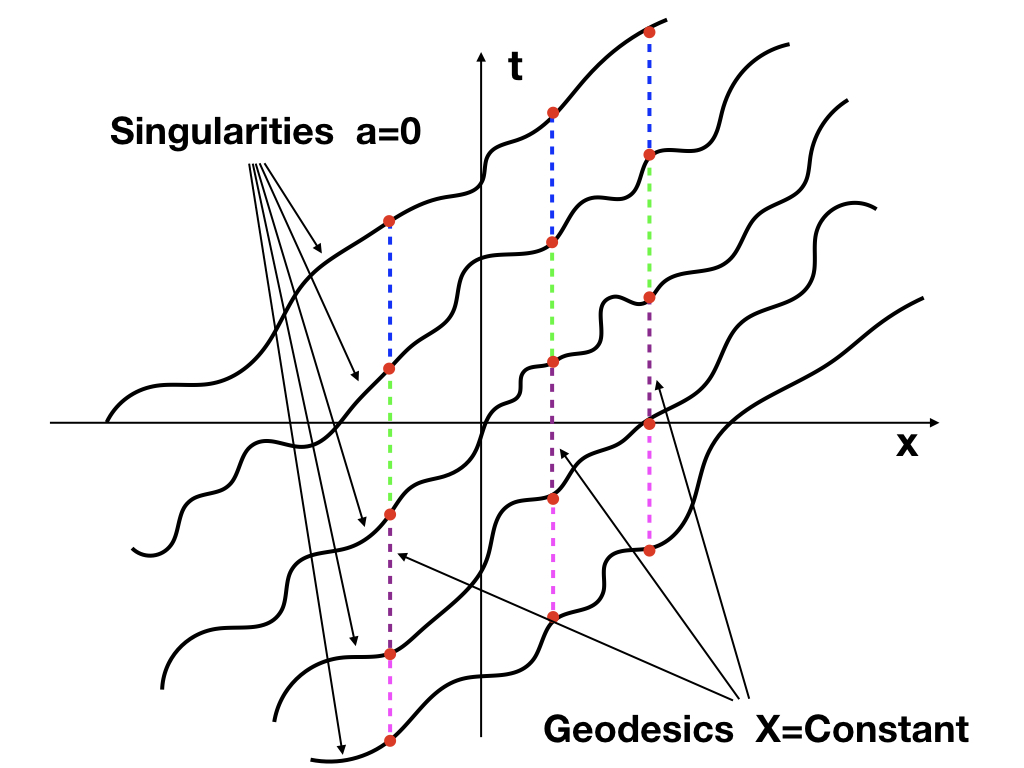}
\includegraphics[scale=0.24]{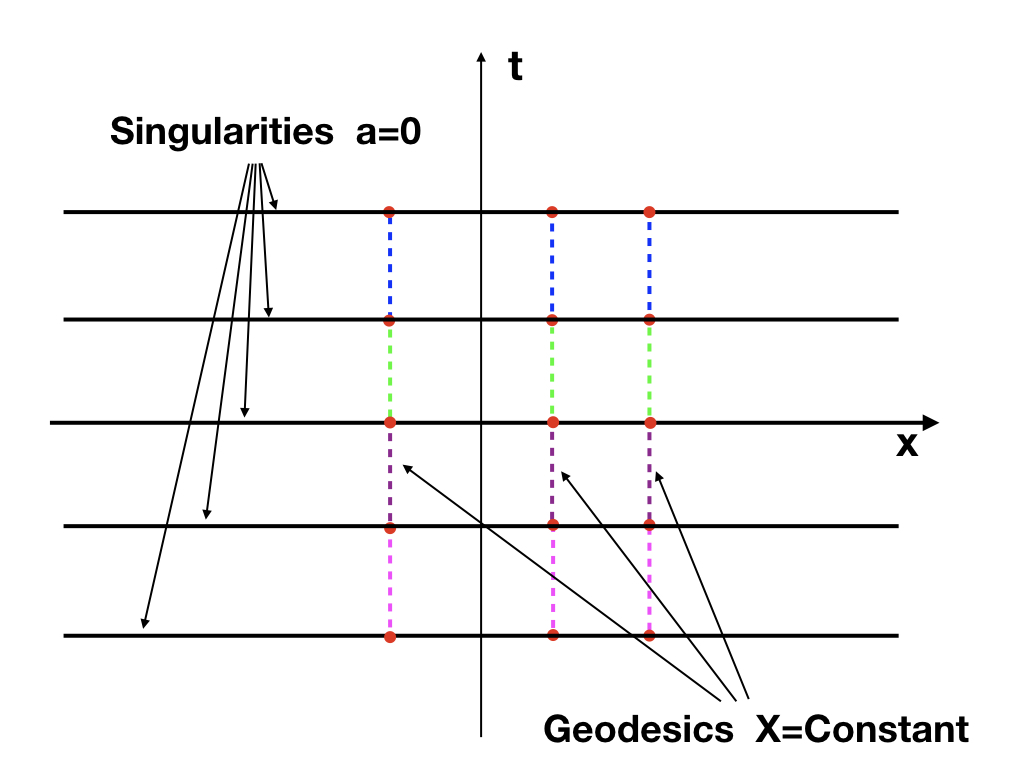}
\caption{\label{singularity}Top: ``micro-cyclic universes" shown in a synchronous reference frame. The curves $a=0$ are singularities where micro ``big bounces" happen. The world lines of particles at rest relative to the reference system are vertical lines $\mathbf{x}=Constants$. They are incomplete geodesics which end at the singularities. Along each segment of the geodesics between the singularities is a ``micro-universe" which starts with a ``micro-big-bang'' and ends with a ``micro-big-crunch".\\\\
Bottom: homogeneous cyclic universes shown in a synchronous reference frame (FLRW). The horizontal lines $a=0$ are singularities where the ``big bounces" happen. The world lines of particles at rest relative to the reference system are vertical lines $\mathbf{x}=Constants$. They are incomplete geodesics which end at the singularities. The whole space simultaneously starts with a big bang and ends with a big crunch.}
\end{figure}

The divergences of $K$, $\sigma^2$ signal that the turning points $a=0$ at which the space switches from contractions to expansions are actually spacetime singularities. These singularities are very similar to the big bang singularity. The alternatively expanding and contracting picture is similar to the cyclic model (or oscillating model) of the universe in the sense that every point in space is a ``micro-cyclic universe" which is following an eternal series of oscillations. Each ``micro-universe" begins with a ``big bang" and ends with a ``big crunch" and then a ``big bounce" happens which bounces the ``crunch" back to a new ``bang" that the cycle starts over again (see FIG. \ref{singularity}). We are going to discuss the singularities in more detail later in Sec. \ref{singularity issue}.

The shear $\sigma^2(t, \mathbf{x})$ measures the local anisotropy of the spacetime. In fact, the dynamics given by the local scale factor $a(t, \mathbf{x})$ is only a description after averaging over different directions. Statistically, the most commonly happening picture in the wildly fluctuating spacetime is that the space is locally expanding in some directions and contracting in others, with the directions of expansion and contraction constantly changing. An initial sphere in this fluctuating spacetime will quickly distort toward an ellipsoid with principle axes given by the eigenvectors of $\sigma^a_{\,\,b}$, with rate given by the eigenvalues of $\sigma^a_{\,\,b}$ \cite{Wald:1984rg}.

So far, we have analyzed the local dynamics of the scale factor $a$ and the shear $\sigma^2$ at a fixed spatial point $\mathbf{x}$. At each such point $\mathbf{x}$, the space is alternatively oscillating between expansion and contraction in every direction and the phases of the oscillations in different directions are commonly different. The global structure of the spacetime would be these small local structures ``glued'' together.

Since on the initial Cauchy surface $\Sigma_0$, $K>0$ and $K<0$ are equally possible initial data, we have that in general the initial conditions $a(0, \mathbf{x})$ and $\dot{a}(0, \mathbf{x})$ for the oscillator equation \eqref{a evolution no vacuum} would take different values at different spatial points. So the phases of these oscillations of $a(t, \mathbf{x})$ at different spatial points would be different. In other words, at any instant of time $t$, the space would be expanding at one point and contracting at neighboring points and vice versa. These phase differences result in a large cancellation between the local expansions and contractions when performing the macroscopic average over the hypersurfaces $\Sigma_t$. The macroscopic average does not require a very large volume: a cubic centimeter contains some $10^{100}$ Planck-size regions. Therefore, we have the average $\langle K\rangle$ over $\Sigma_t$ approaches $0$ for any sensible macroscopic average procedure. The observed macroscopic volume of the space would then approach a constant:
\begin{equation}
V=\int d^3x\sqrt{h}=\int d^3x|a|^3=Constant.
\end{equation}
Thus in this spacetime the large cosmological constant $\lambda_{\mathrm{eff}}$ has huge effect at small scale but becomes hidden macroscopically. This resolves the ``old" cosmological constant problem of explaining why the large vacuum energy does not have large observable gravitational effect. 


\section{The effect of vacuum stress energy tensor fluctuation}\label{The effect of vacuum stress energy tensor fluctuation}
In the last section the vacuum stress energy tensor fluctuation term $F$ is excluded. We have shown that the huge gravitational effect of the expectation value of the large vacuum energy can be hidden by small (Planck) scale spacetime fluctuations when $\lambda'_{\mathrm{eff}}$ takes large negative values. It addresses the old cosmological constant problem but does not explain the observed accelerating expansion of the Universe.

In this section we study the effect of the fluctuation term $F$ on the spacetime dynamics and show that it can serve as the ``dark energy" to accelerate the expansion of the Universe.

$F$ is a linear combination of the components of the vacuum stress energy tensor. It receives contributions from all known and unknown fundamental fields. We are going to use a free massive scalar field $\phi$ as an example to illustrate the key fluctuation properties of $F$ relevant to the dynamics of the system.

In principle, $\phi$ should be treated as a quantum operator $\hat{\phi}$. However, as explained in Sec. \ref{approach} that we do not have a satisfactory quantum theory of gravity yet. For this reason, we are still using the classical Einstein equation \eqref{classical EFT} in which both the metric and the matter fields are classical to study how matter fields vacuum fluctuations affect the spacetime dynamics. In order to do this, we are going to model the quantum field $\hat{\phi}$ as a classical fluctuating field $\phi$ to simulate the quantum fluctuations of $F$.

At each spatial point $\mathbf{x}$, quantum fields can be viewed as an infinite collection of harmonic oscillators. In particular, $\hat{\phi}$ can be expressed as
\begin{eqnarray}\label{field expansion}
&&\hat{\phi}(t,\mathbf{x})\nonumber\\
=&&\int\frac{d^3k}{(2\pi)^{3/2}}\frac{1}{\sqrt{2\omega}}\left(\hat{a}_{\mathbf{k}}e^{-i(\omega t-\mathbf{k}\cdot\mathbf{x})}+\hat{a}_{\mathbf{k}}^{\dag}e^{+i(\omega t-\mathbf{k}\cdot\mathbf{x})}\right)\\
=&&\int\frac{d^3k}{(2\pi)^{3/2}}\left(\hat{x}_{\mathbf{k}}\cos(\omega t-\mathbf{k}\cdot\mathbf{x})+\frac{\hat{p}_{\mathbf{k}}}{\omega}\sin(\omega t-\mathbf{k}\cdot\mathbf{x})\right),\nonumber
\end{eqnarray}
where $\omega=\sqrt{\mathbf{k}^2+m^2}$ and we have used the relations
\begin{equation}
\hat{a}_{\mathbf{k}}=\sqrt{\frac{\omega}{2}}\left(\hat{x}_{\mathbf{k}}+i\frac{\hat{p}_{\mathbf{k}}}{\omega}\right), \quad \hat{a}^\dag_{\mathbf{k}}=\sqrt{\frac{\omega}{2}}\left(\hat{x}_{\mathbf{k}}-i\frac{\hat{p}_{\mathbf{k}}}{\omega}\right)
\end{equation}
to obtain the last line of \eqref{field expansion}. The vacuum state defined by
\begin{equation}\label{vacuum definition}
\hat{a}_{\mathbf{k}}|0\rangle=0, \quad \text{for any}\,\, \mathbf{k},
\end{equation}
is not an eigenstate of the operator coefficients $\hat{x}_{\mathbf{k}}$ and $\hat{p}_{\mathbf{k}}$. The probability densities for $\hat{x}_{\mathbf{k}}$ and $\hat{p}_{\mathbf{k}}$ to take values $x_{\mathbf{k}}$ and $p_{\mathbf{k}}$ are given by the square of their wave functions
\begin{eqnarray}
\langle x_{\mathbf{k}}|0\rangle&=&\left(\frac{\omega}{\pi}\right)^{\frac{1}{4}}e^{-\frac{\omega x_{\mathbf{k}}^2}{2}},\label{wave function 1}\\
\langle p_{\mathbf{k}}|0\rangle&=&\frac{1}{(\pi\omega)^{\frac{1}{4}}}e^{-\frac{p_{\mathbf{k}}^2}{2\omega }}. \label{wave function 2}
\end{eqnarray}

Note that the vacuum state defined by \eqref{vacuum definition} is still Minkowski vacuum, this will be justified in Sec. \ref{define vacuum}.

A natural way to simulate the quantum fluctuations of $\hat{\phi}$ is letting the operator coefficients $\hat{x}_{\mathbf{k}}$ and $\hat{p}_{\mathbf{k}}$ become stochastic constants $x_{\mathbf{k}}$ and $p_{\mathbf{k}}$:
\begin{eqnarray}\label{classical expansion}
&&\phi(t,\mathbf{x})\\
=&&\int\frac{d^3k}{(2\pi)^{3/2}}\left(x_{\mathbf{k}}\cos(\omega t-\mathbf{k}\cdot\mathbf{x})+\frac{p_{\mathbf{k}}}{\omega}\sin(\omega t-\mathbf{k}\cdot\mathbf{x})\right),\nonumber
\end{eqnarray}
where the probability density distributions of $x_{\mathbf{k}}$ and $p_{\mathbf{k}}$ are given by the square of the wave functions \eqref{wave function 1} and \eqref{wave function 2}. This treatment is similar to the Wigner-Weyl description of quantum mechanics we used in \cite{PhysRevD.95.103504}. 

The stress energy tensor $T_{ab}$ is a functional of $\phi$ and $\nabla_a\phi$ defined by
\begin{equation}\label{tab}
T_{ab}=\nabla_a\phi\nabla_b\phi-\frac{1}{2}g_{ab}\left(\nabla_c\phi\nabla^c\phi+m^2\phi^2\right).
\end{equation}
Direct calculation shows that $\rho+\text{tr}T=2\dot{\phi}^2-m^2\phi^2$ and thus from \eqref{F definition} we obtain that the contribution to $F$ from $\phi$ is
\begin{equation}\label{delta rho expression for scalar field}
F=\frac{4\pi G}{3}\left(2\dot{\phi}^2-m^2\phi^2-C\right),
\end{equation}
where $C=\langle 2\dot{\phi}^2-m^2\phi^2\rangle$ is a constant to make sure $\langle F\rangle=0$. Note that the expression \eqref{delta rho expression for scalar field} for $F$ does not explicitly depend the metric $g_{ab}$. 

At each spatial point $\mathbf{x}$, $F$ can be regarded as a time dependent function $F_{\mathbf{x}}(t)$ given by \eqref{delta rho expression for scalar field} and \eqref{classical expansion}. For convenience, we rewrite the key dynamical equation \eqref{a evolution} here as
\begin{equation}\label{full evolution 1}
\ddot{a}+\left(\frac{1}{3}\left(2\sigma^2-\lambda'_{\mathrm{eff}}\right)+F_{\mathbf{x}}(t)\right)a=0.
\end{equation}
The evolution equation for $\sigma^2$ now becomes
\begin{equation}\label{sigma evolution full}
(\sigma^2)^\cdot=-2K\sigma^2-R^{(3)}_{ab}\sigma^{ab}+8\pi G T_{ab}\sigma^{ab}.
\end{equation}

It turns out that the case when $F$ is relatively small compared to $-\lambda'_{\mathrm{eff}}$ is most interesting. In this case, the vacuum stress energy tensor fluctuation serves as a small perturbation of the picture of the micro-cyclic ``universes" we obtained in the last section \ref{The effect of spacetime fluctuation}.

The standard formulation of the cosmological constant problem assumes the effective field theory which is valid only up to some certain high energy cutoff $\Lambda$. We adopt the same assumption in this paper and impose the cutoff $\Lambda$ to the quantum field expansion \eqref{field expansion}. Then the magnitude of the fluctuation of $F$ goes as $\sim G\Lambda^4$ and $F$ is relatively small means
\begin{equation}
-\lambda'_{\mathrm{eff}}\sim-\lambda_B\gg\Lambda^2\geq G\Lambda^4, \quad\text{assuming}\,\, \Lambda\leq E_P,
\end{equation}
where $E_P$ is the Planck energy.

Plugging \eqref{classical expansion} into \eqref{delta rho expression for scalar field} one can obtain a complicated expression for $F_{\mathbf{x}}(t)$ which can be written as the following form:
\begin{equation}
F_{\mathbf{x}}(t)=\int_0^{2\Lambda} d\gamma\left(f_{\mathbf{x}}(\gamma)\cos(\gamma t)+g_{\mathbf{x}}(\gamma)\sin(\gamma t)\right),
\end{equation}
where $f_{\mathbf{x}}$ and $g_{\mathbf{x}}$ are some integrals of $x_{\mathbf{k}}$ and $p_{\mathbf{k}}$ over $\mathbf{k}$.

$F_{\mathbf{x}}(t)$ fluctuates around zero. This fluctuation serves as an external force which changes the parameter $\Omega^2$ of the oscillation system. A dynamical system with time-varying parameters is likely to exhibit parametric resonance phenomenon. A simplest example of parametric resonance is the following harmonic oscillator with periodically perturbed frequency:
\begin{equation}\label{parametric resonance example}
\ddot{x}+\left(\omega_0^2+\epsilon\cos(\gamma t)\right)x=0.
\end{equation}
If $\epsilon=0$, the unperturbed solution to \eqref{parametric resonance example} is simply 
\begin{equation}\label{unperturbed solution}
x(t)=A\cos(\omega_0t+\theta),
\end{equation}
where $A, \theta$ are integration constants which are determined by the initial values $x(0), \dot{x}(0)$. When $\epsilon\ll\omega_0^2$ is small but nonzero, the parametric resonance would happen if the perturbation frequency $\gamma$ closes to $2\omega_0/n$, where $n$ is a positive integer, and the solution perturbed from \eqref{unperturbed solution} becomes unstable. In this case, the amplitude of the oscillation grows exponentially that the perturbed unstable solution is asymptotic to
\begin{equation}
x(t)\sim e^{st}A\cos\left(\omega_0 t+\theta\right), \quad s>1.
\end{equation}
The strength of the parametric resonance characterized by the exponent $s$ decreases as $n$ increases, i.e., as the perturbation frequency $\gamma$ becomes small compared to the oscillator $x$'s natural frequency $\omega_0$. This is easy to understand since as $n\to\infty$, $\gamma\to 0$ so that \eqref{parametric resonance example} reduces to an ordinary harmonic oscillator with constant frequency which has no parametric resonance behavior.

Compared to \eqref{parametric resonance example}, \eqref{full evolution 1} is more complicated in two aspects: i) the external perturbation term $\epsilon\cos(\gamma t)$ in \eqref{parametric resonance example} is periodic which contains only one frequency $\gamma$ while the corresponding term $F_{\mathbf{x}}(t)$ in \eqref{full evolution 1} is not strictly periodic which contains a continuous spectrum of frequencies between $0$ to $2\Lambda$; ii) the natural frequency term $\omega_0^2$ in \eqref{parametric resonance example} is constant while the corresponding term $\frac{1}{3}\left(2\sigma^2-\lambda'_{\mathrm{eff}}\right)$ in \eqref{full evolution 1} is not constant due to the varying shear $\sigma^2$ whose evolution follows \eqref{sigma evolution full}.

Although there are the above two differences, we argue that the dynamical evolution of \eqref{full evolution 1} would exhibit similar parametric resonance phenomenon.

For simplicity, we first ignore the second difference by studying the following simpler equation in which the shear term $\sigma^2$ in \eqref{full evolution 1} has been dropped:
\begin{equation}\label{simpler equation}
\ddot{a}+\left(-\frac{\lambda'_{\mathrm{eff}}}{3}+F_{\mathbf{x}}(t)\right)a=0.
\end{equation}
Similar to \eqref{parametric resonance example} that if we set $F_{\mathbf{x}}(t)=0$ in \eqref{simpler equation}, the solution is
\begin{equation}
a(t, \mathbf{x})=A_{\mathbf{x}}\cos\left(\sqrt{-\frac{\lambda'_{\mathrm{eff}}}{3}}t+\theta_\mathbf{x}\right),
\end{equation}
where $A_\mathbf{x}, \theta_\mathbf{x}$ are integration constants which are determined by the initial values $a(0, \mathbf{x}), \dot{a}(0, \mathbf{x})$.

\begin{figure}
\includegraphics[scale=0.3]{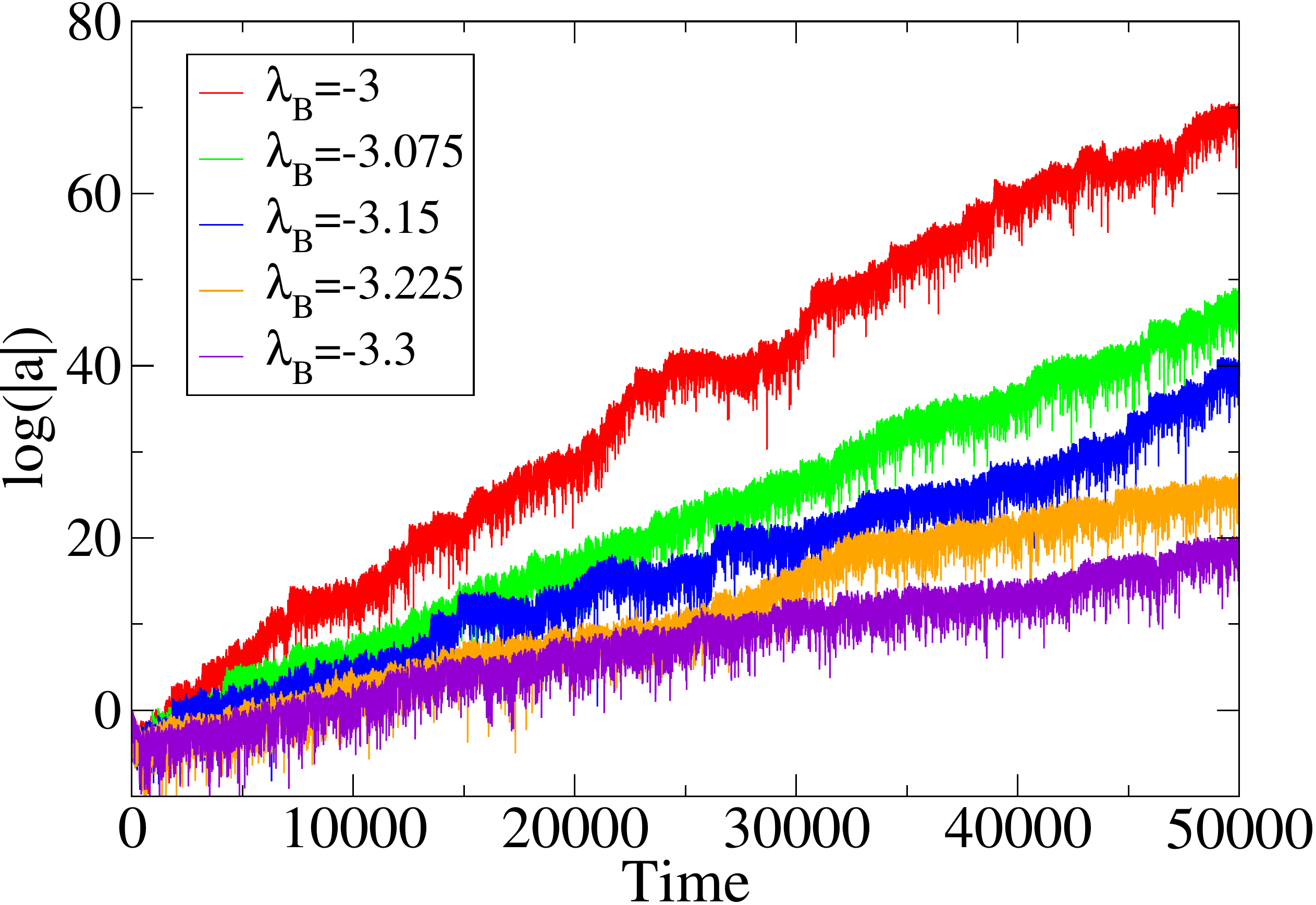}
\caption{\label{new figure3}Numerical simulation for the dependence of $\log|a|$ on the bare cosmological constant $\lambda_B$. We take the cutoff $\Lambda=1$. $400$ samples are averaged for each line. Planck units are used for convenience. The matter fields are one Boson field and one Fermion field. The magnitude of $\langle\rho+\mathrm{tr}T\rangle$ for both fields are set equal but with opposite sign. It shows that the Hubble expansion rate decreases as $-\lambda_B$ increases. We use the same numerical method described in \cite{PhysRevD.95.103504}. (This numerical simulation actually comes from Chapter 10.3 of the thesis \cite{Wang_2018}. The original idea of taking the bare cosmological constant to large negative values actually started in \cite{Wang_2018}. We abandoned the idea there because the problem of large spatial curvature explained in Chapter 10.5 of \cite{Wang_2018}. Fortunately, we found later that this is not a problem in the general approach we used in this paper. It was a problem in \cite{Wang_2018} because we pre-assumed that the metric took the form of Eq.(10.33), which is not true for the general fluctuating spacetime metric \eqref{most general metric} we are studying in this paper. In fact, we have shown in Sec.\ref{The spatial curvature and the cosmological constant} of this paper that in our fluctuating spacetime the bare cosmological constant has to take large negative values to make the observed macroscopic spatial curvature small.)} 
\end{figure}

The occurrence of the parametric resonance does not require a strictly periodic perturbation on the parameter $\Omega^2$. The natural frequency of \eqref{simpler equation} is $\Omega_0=\sqrt{-\lambda'_{\mathrm{eff}}/3}$. There always exists an integer $n_0$ such that for any $n\geq n_0$, $F_{\mathbf{x}}(t)$ contains the frequencies $2\Omega_0/n\in\left(0, 2\Lambda\right)$ which may excite resonances. So the parametric resonance should always happen and the perturbed solution to \eqref{simpler equation} is asymptotic to
\begin{equation}
a(t, \mathbf{x})\sim e^{Ht}A_{\mathbf{x}}\cos\left(\sqrt{-\frac{\lambda'_{\mathrm{eff}}}{3}}t+\theta_\mathbf{x}\right),
\end{equation}
where $H>0$ characterize the strength of the parametric resonance. The straight lines with positive slope in FIG. \ref{new figure3} of the numerical simulation of \eqref{simpler equation} show that the parametric resonance does happen. The exponent $H\to 0$ as $-\lambda_B\to+\infty$ since the relative magnitude of the perturbation term $F_{\mathbf{x}}(t)$ to $-\lambda'_{\mathrm{eff}}/3$ decreases to zero. This property is also shown in FIG. \ref{new figure3} by the decreases of the slopes of the straight lines as the value of $-\lambda_B$ increases.

We can use the same method we used in \cite{PhysRevD.95.103504} to estimate more accurately how $H$ depends on $\lambda_B$ and $\Lambda$. Notice that the small perturbation $F_{\mathbf{x}}(t)$ is also adiabatic since the the time scale of variations of $F_{\mathbf{x}}(t)$ is $t\sim 1/\Lambda$, which is much smaller than $a$'s oscillation period $T\sim 1/\sqrt{-\lambda_B}$. During each period of oscillation of $a$, the frequency $\Omega^2=-\frac{\lambda'_{\mathrm{eff}}}{3}+F_{\mathbf{x}}(t)$ almost does not change so that this is an adiabatic process\footnote{The time scale of variations of $F_{\mathbf{x}}(t)$ can be obtained from the time-energy uncertainty relation
\begin{equation}
\Delta E \Delta t\sim 1.
\end{equation}
The energy scale of the quantum matter fields is just the cutoff scale $\Lambda$. As the change in energy is significant, i.e. $\Delta E\sim\Lambda$, we have $\Delta t\sim 1/\Lambda$. This means $F_{\mathbf{x}}(t)$ would become appreciably different after a time interval of the order $1/\Lambda$. This is easy to understand since the dominant contribution to the stress-enenegy tensor comes from field modes of high frequencies close to $\Lambda$. This result can also be obtained by calculating the correlation functions of the stress energy tensor (see \cite{PhysRevD.95.103504} for a direct calculation).}. Then follow the same steps of Section.VC of \cite{PhysRevD.95.103504} with $\Omega\sim\sqrt{G}\Lambda^2$ replaced by $\Omega\sim\sqrt{-\lambda_B/3}$, we obtain that
\begin{equation}\label{dependence of H on Lambda}
H=\alpha\Lambda e^{-\beta\frac{\sqrt{-\lambda_B}}{\Lambda}},
\end{equation}
where $\alpha, \beta>0$ are two dimensionless constants whose values depend on the detailed fluctuation property of $F_{\mathbf{x}}(t)$.

\begin{figure}
\includegraphics[scale=0.3]{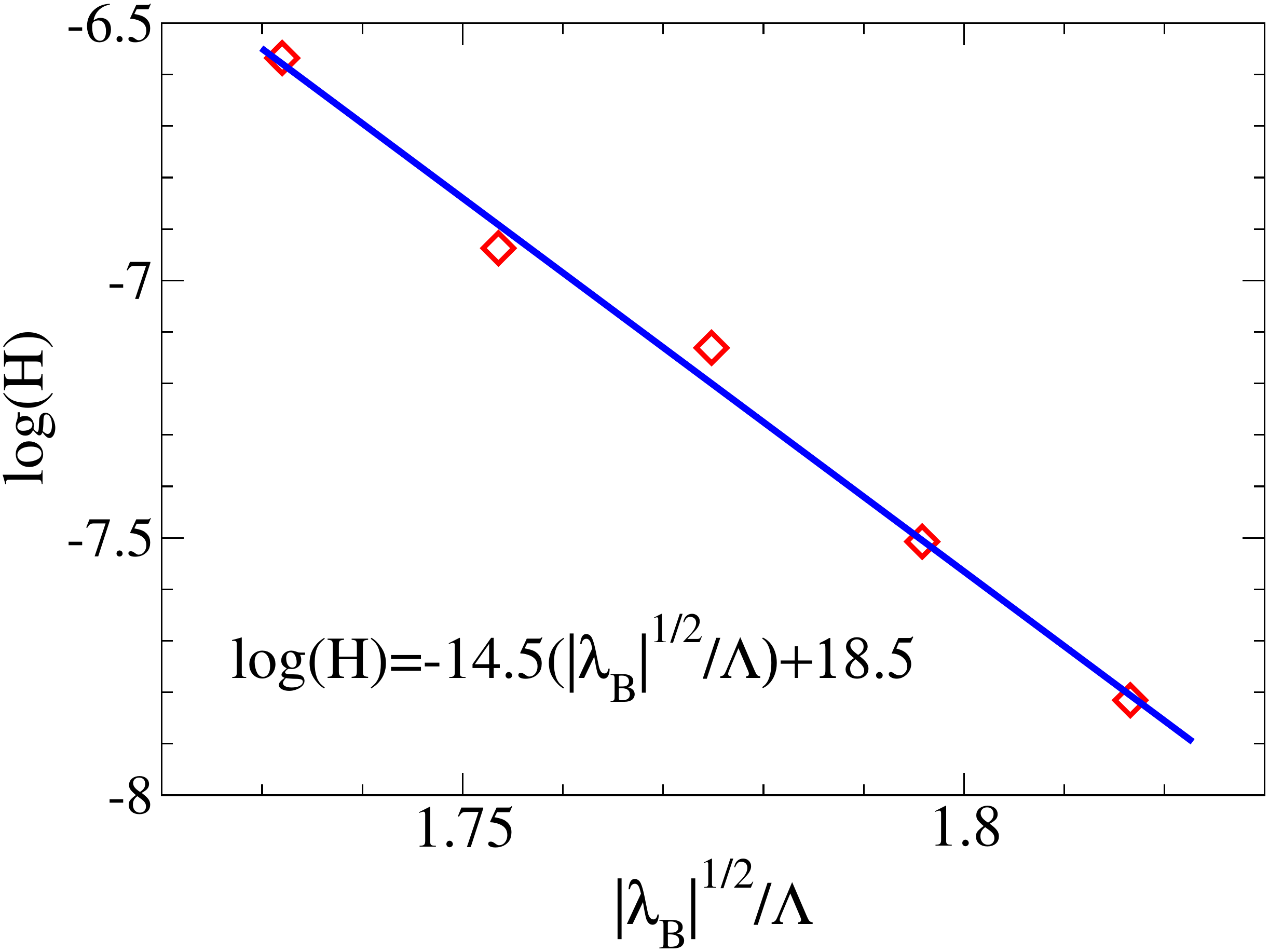}
\caption{\label{new figure4}Plot of $\log H$ over $\sqrt{|\lambda_B|}$ when the matter fields are one Boson field and one Fermion field. The fitting result shows that $\alpha\sim e^{18}, \beta\sim 14$. Planck units are used for convenience. The cutoff $\Lambda=1$.}
\end{figure}

The result \eqref{dependence of H on Lambda} is easy to understand. Larger $\frac{\sqrt{-\lambda_B}}{\Lambda}$ means smaller and slower fluctuations of the perturbation $F_{\mathbf{x}}(t)$ compared to the oscillations of the system and thus a smaller rate of change $H$. The extra factor $\Lambda$ in front of $e^{-\beta\frac{\sqrt{-\lambda_B}}{\Lambda}}$ is because faster fluctuations gives stronger parametric resonance. The fitting result FIG. \ref{new figure4} gives an estimation that $\alpha\sim e^{18}, \beta\sim 14$ if the matter fields are one Boson field and one Fermion field.

Now we put the shear term $\sigma^2$ back to the evolution equation \eqref{full evolution 1}. The dynamics of \eqref{full evolution 1} is way more complicated than \eqref{simpler equation}. However, the parametric resonance should still occur. The root cause of the parametric resonance is that, for certain frequencies of the external perturbations on the parameter $\Omega^2$, the restoring force does more positive work as the oscillator moves toward the equilibrium point than negative work as the oscillator moves away from the equilibrium point. During each cycle of the oscillation, the energy transferred to the system is proportional to the oscillation amplitude. This leads to the exponential growth of the amplitude. After putting the $\sigma^2$ term back, the oscillation of $a$ is no longer sinusoidal and the frequency of the oscillation becomes larger. The exact frequencies which may excite the resonances are no longer the same as \eqref{simpler equation}. But since $F_{\mathbf{x}}(t)$ contains a continuous spectrum of frequencies between $0$ to $2\Lambda$, there should always exist new resonance frequencies lie between $0$ to $2\Lambda$. For this reason, we argue that the parametric resonance still always occur and the perturbed solution to \eqref{full evolution 1} is asymptotic to
\begin{equation}
a(t, \mathbf{x})\sim e^{Ht}a_0(t, \mathbf{x}),\quad H>0,
\end{equation}
where $a_0(t, \mathbf{x})$ is the solution to \eqref{a evolution no vacuum} where the vacuum stress energy tensor fluctuation is excluded. Then the observed macroscopic volume of the space would be
\begin{equation}
V(t)=\int d^3|a|^3=e^{3Ht}V(0).
\end{equation}
So $H$ represents the Hubble expansion rate produced by the vacuum stress energy tensor fluctuations. This produces an accelerated expanding universe. Moreover, we would have
\begin{equation}
H\to 0, \quad \text{as}\,\,\, -\lambda_B\to+\infty,
\end{equation}
since the relative magnitude of the perturbation term $F_{\mathbf{x}}(t)$ to the term $\frac{1}{3}\left(2\sigma^2-\lambda'_{\mathrm{eff}}\right)$ decreases as $\lambda_B$ increases. Therefore, for any cutoff value of $\Lambda$, there is always some value for $\lambda_B$ to match the observed small $H$. This suggests that the vacuum stress energy tensor fluctuation serves as the ``dark energy" which is accelerating the expansion of our Universe. 

Since the basic underlying physical mechanism of the parametric resonance is the same, the dependence of $H$ on $\lambda_B$ and $\Lambda$ should also take the form of \eqref{dependence of H on Lambda}, but with different values of $\alpha$ and $\beta$. Then instead of the usual relation \eqref{contributions to effective cc new new} between the effective cosmological constant $\lambda_{\mathrm{eff}}$ and the bare cosmological constant $\lambda_B$, \eqref{dependence of H on Lambda} gives a new relation:
\begin{equation}\label{lambda dependence on bare lambda}
\lambda^{(\mathrm{new})}_{\mathrm{eff}}=3H^2=3\alpha^2\Lambda^2 e^{-2\beta\frac{\sqrt{-\lambda_B}}{\Lambda}}.
\end{equation}
This relation can be rewritten as
\begin{equation}\label{no fine tuning numeric}
-\frac{\Lambda}{2\beta}\log(\lambda^{(\mathrm{new})}_{\mathrm{eff}})=\sqrt{-\lambda_B}-\frac{\Lambda}{2\beta}\log(3\alpha^2\Lambda^2).
\end{equation}

The numerical simulation shown in FIG. \ref{new figure3} and FIG. \ref{new figure4} gives an estimation that $\alpha$ is somewhere between $e^{10}$ to $e^{20}$, $\beta$ is somewhere between $10$ to $20$. Then if we take $\Lambda=1$ (for convenience, we use Planck units here), we have
\begin{equation}
-\frac{\Lambda}{2\beta}\log(\lambda^{(\mathrm{new})}_{\mathrm{eff}})\sim 10, \quad \frac{\Lambda}{2\beta}\log(3\alpha^2\Lambda^2)\sim 1.
\end{equation}
In this case, since the above two terms are only different by $1$ order of magnitude, the term $\sqrt{-\lambda_B}$ only needs to be tuned to an accuracy of $10^{-1}$ or $\lambda_B$ only needs to be tuned to an accuracy of $10^{-2}$ to satisfy \eqref{no fine tuning numeric}.

In general, the difference in the order of magnitude between the two terms $-\frac{\Lambda}{2\beta}\log(\lambda^{(\mathrm{new})}_{\mathrm{eff}})$ and $\frac{\Lambda}{2\beta}\log(\alpha^2\Lambda^2)$ in \eqref{no fine tuning numeric} is determined by the value of $\Lambda, \alpha$ and $\beta$. $\Lambda$ can take any reasonable value smaller than $1$. The values of $\alpha$ and $\beta$ are determined by the detailed properties of quantum vacuum fluctuations. Basically, for fixed $\Lambda$, the difference becomes smaller if $\alpha$ decreases and $\beta$ increases. Because of the exponential suppression, the extreme fine-tuning of the bare cosmological constant $\lambda_B$ to match the observation is not needed.

\section{Discussions}\label{discussions}
\subsection{The issue of the definition of vacuum state}\label{define vacuum}
Defining the vacuum is actually not trivial in curved spacetime. In Minkowski spacetime, the vacuum state is uniquely defined as the state with lowest possible energy. However, there is no well defined vacuum state in a general curved spacetime. This is already an issue in the standard formulation of the cosmological constant problem, although it is rarely mentioned in the literature\footnote{It is mentioned, for example, in the review article \cite{pittphilsci398}}. The spacetime we are dealing with in this paper is sourced by the bare cosmological constant $\lambda_B$ and the matter fields vacuum stress-energy tensor. This spacetime is wildly fluctuating like Wheeler's spacetime foam that no vacuum state definition in the usual sense is possible.

However, we can still define a state which is ``effectively'' a Minkowski vacuum state below the high energy cutoff $\Lambda$. The spacetime we are interested in is dominated by the bare cosmological constant $\lambda_B$. Its fluctuation happens at the length scale $1/\sqrt{-\lambda_B}$, which is much smaller than the length scale $1/\Lambda$ of the field modes. Then the similar argument we made in \cite{PhysRevD.95.103504} also applies here that the corrections to the field modes with frequencies below $\Lambda$ would be small. In other words, the spacetime should still looks like Minkowski for low frequency field modes. Long wavelength fields ride over the Wheeler's foam seeing only their average properties. This is similar to the behavior of very long wavelength water waves which do not notice the rapidly fluctuating atomic soup over which they ride. In this paper we adopt the effective field theory philosophy that the field theory for matter is valid up to the cutoff $\Lambda$. The long wavelength modes average out the wild fluctuations on the scale of $1/\sqrt{-\lambda_B}$, making them behave like modes in flat spacetime. Therefore, below $\Lambda$, we can approximately define the vacuum state as the lowest energy state as usually done in the ordinary quantum field theory in Minkowski spacetime. In other words, below the cutoff $\Lambda$, the vacuum state we are using in this paper is approximately the usual vacuum state defined in Minkowski spacetime. This justifies the use of the Minkowski vacuum defined by \eqref{vacuum definition}.

\subsection{The validity of the classical treatment of the spacetime evolution}\label{The validity of the classical treatment of the spacetime evolution}
It is usually believed that the Planck length is the scale at which quantum gravitational effects is strong. Thus the classical description of spacetime is supposed to become invalid when one goes to higher than Planck energy scale. The reader may have the concern that this would invalidate our classical treatment of the spacetime evolution which is based on the unquantized Einstein equations \eqref{classical EFT}.

However, the energy scale in our model does not necessarily reach Planck scale. In fact, it is more likely that the energy scale is below Planck. 

Note that there are two parameters, the matter fields cutoff $\Lambda$ and the bare cosmological constant $\lambda_B$ in our new relation \eqref{lambda dependence on bare lambda}. Depends on the value of the two dimensionless constants $\alpha$ and $\beta$, the gravity oscillation scale $\sqrt{-\lambda_B}$ might need to be larger than $\Lambda$ for one or two orders of magnitude. The value of the energy scale $\Lambda$ up to which the effective field theory is valid is not known. The particle physics experiments so far have only tested the field theory up to $Tev$ scale. Provided the huge gap between the energy scale of the standard model of particle physics ($10^3 GeV$) and the Planck scale ($10^{19}GeV$), there is actually large chances for $\Lambda$ to take values far below the Planck energy.

For example, it is estimated in \cite{PhysRevD.65.025006} that an upper limit on the domain of validity of the quantum field theory description of nature is around $100\, TeV (10^{-14}E_P)$. If so, the oscillation scale of the gravity field $\sqrt{-\lambda_B}$ could be around $10^{-13} E_P$ or $10^{-12} E_P$, far below the Planck energy and classical general relativity is expected to be a valid description of gravity.

As far as we know, the energy scale of most field theories beyond the standard model is below Planck energy scale. Probably the one most close to the Planck energy is the Grand Unified Theory. If $\Lambda$ is taken to be on the GUT scale which is around $10^{-3} E_P$, then $\sqrt{-\lambda_B}$ could be around $10^{-1} E_P$ or $10^{-2} E_P$. In this case, our classical treatment of gravity should still be valid.

Even if the quantum field theory description of matters is indeed valid until the Planck scale, i.e., if we let $\Lambda=E_P$, then the gravity oscillation scale $\sqrt{-\lambda_B}$ might need to be $10 E_P$ or $100 E_P$ (If we use the values of $\alpha$ and $\beta$ given by the numerical simulation shown in FIG. \ref{new figure3} and FIG. \ref{new figure4}, $\sqrt{-\lambda_B}$ would be around $10 E_P$.). In this case, our classical treatment may not be a precise description of the spacetime evolution. However, since this is not too far above the Planck energy, and, since one of the most important features of a quantum system is the quantum fluctuation arose from the uncertainty principle, our classically fluctuating spacetime should, at least to a certain extent, reveal some quantum fluctuation feature of the future satisfactory quantum theory of gravity.

In fact, this is one of the key points of this paper---the quantum gravity fluctuations at small (Planck) scale are important, it causes a highly inhomogeneous spacetime which can hide the large gravitational effect of quantum vacuum at that scale. Although we do not have a quantum theory of gravity to precisely describe it, we may use classically fluctuating spacetime to approximate it. Note that the standard formulation of the cosmological constant problem also treats the spacetime as classical. It missed the important spacetime fluctuations. In this paper we follow the same classical treatment but include the effect of these fluctuations. The result shows that we may not need to wait until a completely satisfactory theory of quantum gravity to solve the cosmological constant problem and the solution presented in this paper could provide a hint about what the final quantum gravity theory looks like. Our result suggests that such a theory may exhibit a similar micro-cyclic ``universes" picture.

Our classical treatment of the spacetime evolution might also become invalid at the singularities $a=0$. The existence of the singularity is a common issue of classical general relativity, not a particular issue of our model. We are going to discuss this issue in the next subsection.

\subsection{The issue of the singularities}\label{singularity issue}
Probably the biggest concern about this proposal for addressing the cosmological constant problem is the appearance of the singularities at $a=0$.

The existence of singularities is a generic feature of the solution of Einstein field equations under rather general energy conditions (e.g. strong, weak, dominant etc.), which is guaranteed by Penrose-Hawking singularity theorems \cite{PhysRevLett.14.57, Hawking511, Hawking490, Hawking187, Hawking:1973uf, Hawking:1969sw}. Our negative cosmological constant dominated spacetime satisfies the strong energy condition and thus the occurrence of the singularities is inevitable. In fact, it has been shown in \cite{1976ApJ...209...12T} that all timelike geodesics in a globally hyperbolic spacetime dominated by a negative cosmological constant are future and past incomplete\footnote{The anti-de Sitter space is geodesically complete, but it is not globally hyperbolic. All physically realistic spacetimes should be globally hyperbolic.}, and no timelike curve has a proper time length greater than $\pi\sqrt{-3/\lambda_B}$ (see FIG. \ref{singularity}). One can also show that they are curvature singularities since the Kretschmann invariant $R_{abcd}R^{abcd}$ is divergent.

At the singularities $a=0$ the frequency $\Omega^2=+\infty$ since the shear scalar $\sigma^2$ diverges there. This leads to the divergence of the the ``velocity" $\dot{a}$ and the ``acceleration'' $\ddot{a}$ at $a=0$. Physically, this implies that the oscillator $a$ would go across its equilibrium point infinitely fast so that a singular bounce happens within infinitely short of time. The singularities form and then disappear immediately. However, mathematically, there is ambiguity to continue the solution across the singularities. Due to the divergences of $\dot{a}$ at $a=0$, one cannot tell how the solution for $a$ before the crossing over $0$ connected to the solution after the crossing. One basic question raised by this issue is whether the bounces are ``elastic", i.e., whether the magnitude of $\dot{a}$ ``right before" and ``right after" the crossing of $a=0$ equal?

This ambiguity represents the break down of the classical description of gravitation at the singularities and general relativity ``partially" loses its predictability there. We use the wording ``partially" because the classical Einstein equation does not completely lose its predictability at $a=0$, at least follow the classical evolution equation \eqref{a evolution} one can obtain that $a$ must pass $0$ without stopping so that a bounce must happen, although one can not determine unambiguously whether the bounce is ``elastic" or not. In principle, quantum effect of gravity itself would play a dominant role near the singularities so that one need to use quantum gravity to predict what is really going on when $a$ approaches the singularities. Unfortunately, there is no satisfactory quantum theory of gravity yet. However, a natural guess from the ``energy conservation" consideration is that the bounces should be ``elastic", although in general there is no well defined energy for gravitational field in general relativity.

In the following we argue that a natural classical prescription to extend the Einstein field equations beyond the singularities do predict the ``elastic" bounces.

One essential feature of the singularities happened in our picture is that the determinant $g=-a^6$ of the metric becomes $0$ when $a=0$, i.e., the metric becomes degenerate. The standard formulation of general relativity does not allow the metric to be degenerate because the inverse metric $g^{ab}$ would become singular and the quantities involved in Einstein equations like $R^{a}_{bcd}$, $R_{ab}$, $R$ would take on the form $0/0$. A natural way of resolving this kind of singularities characterized by the the vanishing of $g$ is by multiplying both sides of the Einstein equations by some power of $g$:
\begin{equation}\label{densitized einstein equation}
(-g)^pG_{\mu\nu}+(-g)^p\lambda_Bg_{\mu\nu}=(-g)^p8\pi GT_{\mu\nu}.
\end{equation}
For suitable values of $p$, there is no longer any denominator in the above equation \eqref{densitized einstein equation}. This equation is equivalent to the original Einstein equation at points away from the singularities and still valid at the singularities if the metric components $g_{\mu\nu}$ are smooth (or at least their first two derivatives exist) at $a=0$. Then there is no problem for $g_{\mu\nu}$ to unambiguously (uniquely) evolve across the singularities according to the extended equation \eqref{densitized einstein equation}.

This idea of resolving a singularity by mulptiplying Einstein equations with some power of the determinant of the metric is not new. Einstein himself had proposed this idea with his collaborator Rosen in 1935 (for which they credited this idea to Mayer) \cite{PhysRev.48.73} in the study of spacetime metric which is called the Einstein-Rosen bridge later. Ashtekar used a similar trick in his method of ``new variables'' to develop an equivalent Hamiltonian formulation of general relativity \cite{PhysRevD.36.1587}. Stoica followed this idea and formulated the ``singular general relativity" \cite{Stoica:2013wx} which allows the metric to become degenerate. In this fomulation, he argues that not tensor but tensor densities are the physical quanties and the densitized Einstein equations \eqref{densitized einstein equation} are actually more fundamental than the usual Einstein equations \cite{Stoica:2013wx, doi:10.1142/S0219887814500418, Stoica:2012my, Stoica:2012gb, Stoica:2011xf, Stoica:2011nm, Stoica:2014tpa, Stoica:2015yfa, Stoica:2015hba}.

Unfortunately, $g_{\mu\nu}$ is not smooth at the singularities in our spacetime since $g_{\mu\nu}$ itself and its first two derivatives may be divergent at $a=0$. Because of this, even the extended equation \eqref{densitized einstein equation} is not valid at the singularities. However, this issue may be fixed by operating on the metric density $(-g)^p g_{\mu\nu}$ (of suitable weight $-2p$) instead of metric $g_{\mu\nu}$ in \eqref{densitized einstein equation}. The metric density $(-g)^p g_{\mu\nu}$ and its first two derivatives can always be made finite for suitable values of $p$ due to the vanishing of $g$ at the singularities. Then if we replace the argument $g_{\mu\nu}$ by $(-g)^p g_{\mu\nu}$ in \eqref{densitized einstein equation}, i.e., if we express $G_{\mu\nu}$ in terms of $(-g)^p g_{\mu\nu}$ instead of $g_{\mu\nu}$, \eqref{densitized einstein equation} would be valid at the singularities. Then with the requirement that $(-g)^p g_{\mu\nu}$ are smooth (or at least their first two derivatives exist) at the singularities $a=0$ for all suitable values of $p$, the new variables $(-g)^p g_{\mu\nu}$ can unambiguously (uniquely) evolve across the singularities according to the extended equation \eqref{densitized einstein equation}. We can then obtain the solution for the metric $g_{\mu\nu}$ from $(-g)^p g_{\mu\nu}$. In this sense, one can still predict how $g_{\mu\nu}$ evolves beyond the singularities according to the extended Einstein equations \eqref{densitized einstein equation}, although $g_{\mu\nu}$ may still be divergent at the singularities. The divergence of $g_{\mu\nu}$ might not as bad as usually thought since practically one can not measure physical quantities at a point. Any measurement has to be made in a finite region of spacetime and thus there is always an integral $\int d^4x\sqrt{-g}$ which may cancel (or at least weaken) the divergence at the singularity because of the vanishing of $g$ there.

For example, we can apply the above prescription of singularity resolution to the simple Kasner metric
\begin{equation}\label{kasner}
ds^2=-dt^2+t^{2p_1}dx^2+t^{2p_2}dy^2+t^{2p_3}dz^2, \quad  t>0,
\end{equation}
with
\begin{equation}
p_1+p_2+p_3=p_1^2+p_2^2+p_3^2=1.
\end{equation}
The Kasner metric for $t<0$ takes the same form
\begin{equation}
ds^2=-dt^2+(t^2)^{p'_1}dx^2+(t^2)^{p'_2}dy^2+(t^2)^{p'_3}dz^2, \quad  t<0,
\end{equation}
with
\begin{equation}
p'_1+p'_2+p'_3={p'_1}^2+{p'_2}^2+{p'_3}^2=1.
\end{equation}

If we arrange $p_1, p_2, p_3$ in the order $p_1<p_2<p_3$, their values will lie in the intervals
\begin{equation}
-\frac{1}{3}\leq p_1\leq0,\, 0\leq p_2\leq\frac{2}{3},\, \frac{2}{3}\leq p_3\leq 1.
\end{equation}

The metric component $t^{2p_1}$ is divergent at $t=0$ and all the first and second derivatives $(t^{2p_i})^{\cdot}$, $(t^{2p_i})^{\cdot\cdot}$, $i=1,2,3$ are also divergent at $t=0$. So the original Einstein equations are invalid at $t=0$ and thus cannot predict how the metric \eqref{kasner} for $t>0$ evolve across $t=0$ to negative values of $t$. In other words, one cannot tell how the Kasner index $p_1, p_2, p_3$ for $t>0$ relate to the Kasner index $p'_1, p'_2, p'_3$ for $t<0$ from the original Einstein equations. 

However, since the determinant $-g=t^2$, then the metric density $(-g)^pt^{2p_i}=t^{2(p+p_i)}$ and their first and second derivatives are all finite at $t=0$ if $p\geq -p_1+1$ and thus the extended Einstein equations \eqref{densitized einstein equation} for the new variables $(-g)^p g_{\mu\nu}$ are valid there.    

In particular, for $p=-p_1+1$, the three components of $(-g)^pt^{2p_i}$ for $t>0$ are
\begin{equation}\label{densitized metric 1}
t^2, \quad t^{2(p_2-p_1+1)}, \quad t^{2(p_3-p_1+1)},
\end{equation}
and the three components of $(-g)^pt^{2p'_i}$ for $t<0$ are
\begin{equation}\label{densitized metric 2}
t^{2(p'_1-p_1+1)}, \quad t^{2(p'_2-p_1+1)}, \quad t^{2(p'_3-p_1+1)}.
\end{equation}
At $t=0$, the three components of \eqref{densitized metric 1} and their first derivatives are all equal to zero, which is the same as the values of the three components of \eqref{densitized metric 2} and their first derivatives. However, the second derivative of the first component $t^2$ in \eqref{densitized metric 1} at $t=0$ is equal to $2$, while the second derivative of the first component $t^{2(p'_1-p_1+1)}$ in \eqref{densitized metric 2} is $2(p'_1-p_1+1)(2(p'_1-p_1)+1)t^{2(p'_1-p_1)}$, which is either $0$ or $\infty$ if $p'_1\neq p_1$. In order to match them, one must have
\begin{equation}
p'_1=p_1,\quad p'_2=p_2, \quad p'_3=p_3.
\end{equation}

Therefore, we have obtained that the metric density $t^{2(p+p_i)}$, which is valid for all values of $t$, is a unique solution to the extended Einstein equations \eqref{densitized einstein equation} under the requirement that the new variables $(-g)^p g_{\mu\nu}$ are smooth (or at least first two derivatives match) at the singularities for all suitable values of $p$. Then the Kasner metric \eqref{kasner} can be extended to negative values of $t$:
\begin{eqnarray}\label{kasner whole}
ds^2=-dt^2+&&(t^2)^{p_1}dx^2+(t^2)^{p_2}dy^2+(t^2)^{p_3}dz^2,\\
 &&-\infty<t<+\infty. \nonumber
\end{eqnarray}
This metric describes an ``elastic" singular bounce at $t=0$. The function $(t^2)^{p_i}$ is defined as function composition $t\mapsto t^2 \mapsto (t^2)^{p_i}$, where the exponentiation to the real power $p_i$ of the non-negative base $t^2\geq 0$ is defined by extending the usual rational powers to reals by continuity. In this definition, one always has $(t^2)^{p_i}\geq 0$.

The spacetime evolution in this paper can also be continued across the singularities $a=0$ by applying this prescription. It is natural that the bounces given by this prescription is ``elastic" since the smoothness requirement (or at least the first two derivatives continuous) of the new variables $(-g)^p g_{\mu\nu}$ at the singularities $a=0$.

There are some similarities in the singularity structure between our fluctuating spacetime dominated by the negative bare cosmological constant and the Kasner metric \eqref{kasner whole}. In fact, an exact solution of the Einstein equations for a Bianchi-I universe (homogeneous but anisotropic universe with flat spatial curvature) in the presence of a negative cosmological constant has been shown to be Kasner type \cite{Kamenshchik:2009dt}.

The exact dynamics of the metric near the singularities in our wildly fluctuating spacetime is of course much more complicated than the simple Kasner metric \eqref{kasner whole}. According to Wheeler's insight that ``matter doesn't matter" near a singularity, we have, for most type of matter including the negative cosmological constant, the effect of the matter fields on the dynamics of the geometry becomes negligible near the singularity. And also according to the BKL conjecture \cite{Belinsky:1970ew}, near the singularity the evolution of the geometry at different spatial points decouples, we have the dynamics of our spacetime near the singularities should be similar to the BKL singularity, which is a model of the dynamic evolution of the Universe near the initial singularity, described by an anisotropic, chaotic solutions of the Einstein field equations of gravitation. The difference between the usually studied homogeneous BKL model and our fluctuating spacetime is that in our model the spacetime is inhomogeneous that the singularities happen at different times. Also, unlike the usual study that the singularity is supposed to be an end of BKL dynamics, in our spacetime the singularity is not an end but a bounce.

So far, we have argued that the classical spacetime evolution in this paper can continue beyond the singularities if we operate on the metric density $(-g)^pg_{\mu\nu}$ (of suitable weight $-2p$) instead of on the metric $g_{\mu\nu}$ in the extended Einstein equations \eqref{densitized einstein equation}. The price is that we accept the divergence of $g_{\mu\nu}$ at the singularities. This prescription naturally predicts classical ``elastic" bounces at the singularities.

We emphasize here that this way of singularity resolution is just a prescription for trying to keep a classical description of spacetime near the singularities. In principle, quantum effects of gravity itself would play a dominant role near the singularities which would invalidate the classical description of spacetime. We need a quantum theory of gravity to predict what is really going on near the singularities.

Although there is still no satisfactory quantum theory of gravity yet, it is interesting to notice that some existing quantum gravity theory also predicts the similar bounces predicted by the classical evolution equation \eqref{a evolution}. In fact, loop quantum gravity has obtained similar bounce pictures in FLRW, Bianchi, and Gowdy models \cite{Ashtekar:2008zu, Agullo:2016tjh}. The difference is that in loop qunatum gravity the singularity is avoided since the bounce happens before the singularity forms. That is because in the framework of loop quantum gravity the quantum geometry creates a new repulsive force which is totally negligible at low spacetime curvature but rises very rapidly in the Planck regime which bounces the contraction back to expansion. Unlike in our ``classical" model, the structures in loop quantum cosmology before and after the quantum bounce can change in general. For example, quantum gravitational effects can cause Kasner transitions in Bianchi spacetimes \cite{PhysRevD.86.024034, Wilson_Ewing_2018}. However, these differences in the details of the bouncing dynamics should not alter our main result as long as the (average) bounces are ``elastic".
 
One might also feel strange that the local scale factor $a$ can be negative, which contradicts our impression of positive $a$ in standard cosmological models. However, this is not a problem since the physical quanties are always the non-negative determinant $h=a^6\geq 0$ and the positive-(semi)definite spatial metric $h_{ab}$. Gielen and Turok described a similar picture which they called ``perfect quantum cosmological bounce" in the usual homogeneous FLRW universe \cite{PhysRevD.95.103510, PhysRevLett.117.021301}. They also showed that it is natural to extend the scale factor $a$ to negative values, allowing a large, collapsing universe to evolve across a quantum bounce to an expanding universe. They circumvented the big bang singularity by analytically extending $a$ to the entire complex plane that the universe evolves from large negative $a$ to large positive $a$ along a contour which avoids $a=0$.

Another issue caused by the singularities is how they affect the propagation of quantum fields riding on the spacetime. It has been argued in Sec.VII and Sec.IXB of \cite{PhysRevD.95.103504} by direct calculations for a special toy metric that the singularities do not affect the propagation of low frequency field modes. In that toy model, the singularities do not cause problems at the observable low energy regime. This result should be still valid in our general fluctuating spacetime with the singularities---after all, the singularities only appear (and immediately disappear) at energy scales of $\sqrt{-\lambda_B}$, which should not affect the low energy physics whose energy scale is far below $\sqrt{-\lambda_B}$.

\subsection{Advantages of the new scenario}\label{Advantages of the new scenario}
The old scenario presented in \cite{PhysRevD.97.068301} has a couple of shortcomings:

i) In the old scenario, the high energy cutoff $\Lambda$ violates the usual Lorentz invariance requirement of the quantum vacuum \cite{PhysRevD.97.068301}, which leads to the violation of the usually assumed vacuum equation of state \eqref{vacuum equation of state}(i.e., $\langle P\rangle=-\langle\rho\rangle$).

ii) In the old scenario, we required that the square of the time dependent frequency $\Omega^2=4\pi G(\rho+\sum_{i=1}^{3}P_i)/3>0$ (Eq.(42) in \cite{PhysRevD.95.103504}). It would be a disaster to this scenario if there is any significant chance that $\rho+\sum_{i=1}^{3}P_i$ becomes negative \cite{PhysRevD.97.068301, PhysRevD.97.068302}. However, in principle, $\rho+\sum_{i=1}^{3}P_i$ receives contribution from all fundamental fields. Naive calculations show that Boson fields have positive energy density while Fermion fields have negative energy density \cite{Martin:2012bt}. Since we do not have the knowledge of all fundamental fields, the sign of $\rho+\sum_{i=1}^{3}P_i$ can not be determined.

iii) In the old scenario, we required taking $\Lambda$ to super-Planck scale and the oscillation scale of gravity field would be on super-super-Planck scale. However, general relativity is generally expected to break down at or above Planck scale and QFT may break down even earlier.

The new scenario does not have the above listed shortcomings:

i) The high energy cutoff $\Lambda$ just labels the energy scale which measures the magnitude of the quantum fluctuations. Since the bare cosmological constant $\lambda_B$ is dominant, whether or not the usually assumed vacuum equation of state \eqref{vacuum equation of state} is violated at small (Planck) scale does not matter. The different regularization methods for calculating the expectation value of the vacuum energy density do not alter the general scenario, although it may alter the details that the numerical values of the constants $\alpha, \beta$ in \eqref{dependence of H on Lambda} may change.

ii) For our new model to work, we only need to adjust $-\lambda_B\gg\Lambda^2\geq G\Lambda^4$ (assuming $\Lambda\leq E_P$) to make sure the expectation value $\langle\Omega^2\rangle\gg\Lambda^2$ that even a small probability for $\Omega^2<0$ does not matter.

iii) For our new result \eqref{lambda dependence on bare lambda}, $\Lambda$ can take any possible value below the Planck energy and the oscillation scale of the gravity field, which is given by $\sqrt{-\lambda_B}$, would not be far above the Planck energy scale. More detailed discussion has been presented in Sec.\ref{The validity of the classical treatment of the spacetime evolution}.

\subsection{Open questions}
Here we list some open questions raised by this new scenario which deserve further studies in the future.

i) We model both the metric and the matter fields as classical fluctuating fields. What if we treat both of them as quantum operators? Can we get the same result? In other words, what about quantum gravity? As explained in Sec.\ref{The validity of the classical treatment of the spacetime evolution}, if the cutoff $\Lambda$ is far below Planck energy, the classical treatment should be a good approximation. If $\Lambda$ closes to Planck energy, quantum gravity effect might be important. We have argued that one of the most important features of a quantum system is the quantum fluctuation due to the uncertainty principle. When $\Lambda$ closes to Planck energy, this classical treatment should at least reveal some quantum fluctuation picture of the future satisfactory quantum theory of gravity. So can this guide us to the right way to quantize gravity?

ii) What about low energy Einstein equations? Do they decouple from these high energy equations? If not, are there problems with gravity waves from Ligo, or with Planetary motion?

iii) Can this scenario be extended to inflation? Is inflation extra low energy equations, or is inflation also by the similar mechanism? For example, can the phase transitions in the early universe effectively shift the negative bare cosmological constant $\lambda_B$ to values comparable to the zero point fluctuations that the parametric resonance becomes strong and thus be able to produce the inflation? If so, the advantage of this model is that the inflation can be driven by the fluctuations of quantum vacuum of known physical field, without the need to introduce a hypothetical inflaton field with an artificial slow roll potential.

iv) The singularities are probably the most crucial open question, while we have argued that we can push through them by the prescription described in Sec. \ref{singularity issue} which is still in the classical gravity framework, they remain problematic without a satisfactory quantum theory of gravity since in principle quantum gravitational effect should dominant near the singularities. Loop quantum gravity has obtained similar bounce pictures for simpler models, what if also apply it to this model? Will we still get elastic bounces so that our final result still hold?

v) Our picture requires a large negative bare cosmological constant. Interestingly, this is also found in the asymptotic safety program of quantum gravity when studying the RG flow of gravity coupled to the matter fields of the standard model (see, e.g. \cite{Biemans:2017zca, Bonanno:2017gji, Bonanno:2018gck}). The relation between these results deserve further investigations. They might be helpful to answer some of the above open questions.

vi) Both our model and the anti-de Sitter space has a negative cosmological constant. If ignore the relatively small perturbation produced by vacuum stress energy tensor, this spacetime is called an Einstein manifold with a negative cosmological constant. The difference is that anti-de Sitter space is homogeneous while our model is highly inhomogeneous. Is there any relation between our model and the ADS/CFT correspondence?

\section{Summary and conclusion}\label{Summary and conclusion}
The cosmological constant problem arises from the following two basic principles of quantum mechanics and general relativity:

Principle $1$: the uncertainty principle which requires that quantum fields vacuum has a large energy density $\langle\rho\rangle\sim\Lambda^4$;

Principle $2$: the equivalence principle which predicts that the large energy of quantum vacuum must gravitate to produce large gravitational effect.

It is well known that classical general relativity and quantum mechanics is incompatible with each other and there is not a satisfactory quantum theory of gravity yet to combine them together. So in order to study the large gravitational effect produced by the large  quantum vacuum energy, the quantum vacuum is usually modeled as some classical source of gravity so that one can still apply the classical general relativity.

The standard formulation of the cosmological constant problem models the quantum vacuum as a perfect classical fluid. It assumes the following properties of the classical fluid and the spacetime it rests on:

Assumption $1$: the energy density of the fluid is constant;

Assumption $2$: the spacetime are homogeneous and isotropic so that one can use the FLRW metric \eqref{FLRW}.

Then based on the above two assumptions, the standard formulation obtains that the observed effective cosmological constant $\lambda_{\mathrm{eff}}=3H^2\sim\lambda_B+G\Lambda^4$. In order to cancel the large gravitational effect characterized by the term $G\Lambda^4$, the bare cosmological constant $\lambda_B$ has to be fine-tuned to extreme accuracy to obtain a small $\lambda_{\mathrm{eff}}$.

``Conventional" approaches to tackle this problem are either trying to modify quantum mechanics in some way to make vacuum energy density small or trying to modify general relativity in some way to make vacuum energy not gravitate. Some approaches are even pleading to the anthropic arguments. 

In this paper, we notice that the Assumptions 1 and 2 are not true at small scales. The large vacuum energy density is produced by small scale quantum fluctuations, there is no reason to apply the cosmological scale FLRW metric to the small scale phenomenon. In our ``unconventional" approach, we model the quantum vacuum as a classical fluctuating field and uses the general metric \eqref{new metric} to describe the fluctuations. We assumes the following properties of the classical field and the spacetime fluctuations:

Assumption $1'$: the classical fluctuating field is modeled by \eqref{classical expansion};

Assumption $2'$: the initial data $K_{ij}>0$ and $K_{ij}<0$ on the hypersurface $\Sigma_0$ are equally possible.

Then from the Assumption $2'$ we obtain that the fluctuations of the spacetime would produce a large positive contribution to the averaged macroscopic spatial curvature of the Universe. In order to cancel this contribution to match the observation, the usually defined effective cosmological constant $\lambda_{\mathrm{eff}}$ by \eqref{contributions to effective cc new new} has to take a large negative value. The spacetime dynamics sourced by this large negative $\lambda_{\mathrm{eff}}$ would be similar to the cyclic model of the universe in the sense that at small scales every point in space is a ``micro-cyclic universe" which is following an eternal series of oscillations between expansions and contractions. The turning points $a=0$ at which the space switch from contraction to expansion are curvature singularities, we assumed a prescription in Sec.\ref{singularity issue} to continue the space time evolution beyond the singularities by a natural extension of the Einstein equations at the singularities. Because the phases of the oscillations of the ``micro-cyclic universes" at different spatial points are different, the effect of these oscillations cancel and the large cosmological constant $\lambda_{\mathrm{eff}}$ is screened. These phase differences are primarily produced by the ``active" fluctuations of gravity itself.

When the ``passive" fluctuations of the spacetime induced by the quantum vacuum stress tensor fluctuation are considered and if the bare cosmological constant $\lambda_B$ is dominant over the vacuum stress tensor fluctuation, the size of each ``micro-universe" would increase a tiny bit at a slowly accelerating rate during each micro-cycle of the oscillation due to the weak parametric resonance effect produced by the fluctuations of the quantum vacuum stress energy tensor. We obtain a new relation $\lambda^{\mathrm{new}}_{\mathrm{eff}}=3H^2\sim\Lambda^2 e^{-2\beta\sqrt{-\lambda_B}/\Lambda}$ which shows that the contribution from the vacuum energy density to the accelerating expansion of the Universe is exponentially suppressed. In this way, the large cosmological constant generated at small scales is hidden at observable scale and no fine-tuning of $\lambda_B$ to the accuracy of $10^{-122}$ is needed. This at least resolves the old cosmological constant problem and suggests that it is the quantum vacuum fluctuations serve as the dark energy which is accelerating the expansion of our Universe. This mechanism shows that the physics happens at the smallest (Planck) scale may have effects on the largest (cosmological) scale.

\bibliographystyle{unsrt}
\bibliography{how_vacuum_gravitates}

\end{document}